\documentclass[preprint]{ptephy}

\preprintnumber{KYUSHU-HET-144}

\usepackage{amssymb}
\usepackage{amsthm}
\usepackage{amsmath}
\usepackage{booktabs}

\DeclareMathOperator{\re}{Re}

\numberwithin{equation}{section}

\usepackage{url}

\begin{document}

\title{Renormalizability of the gradient flow in the 2D $O(N)$ non-linear sigma
model}

\author{%
\name{\fname{Hiroki} \surname{Makino}}{1} and
\name{\fname{Hiroshi} \surname{Suzuki}}{1,\ast}
}

\address{%
\affil{1}{Department of Physics, Kyushu University, 6-10-1 Hakozaki, Higashi-ku, Fukuoka, 812-8581, Japan}
\email{hsuzuki@phys.kyushu-u.ac.jp}
}

\begin{abstract}
It is known that the gauge field and its composite operators evolved by the
Yang--Mills gradient flow are ultraviolet (UV) finite without any
multiplicative wave function renormalization. In this paper, we prove that the
gradient flow in the 2D $O(N)$ non-linear sigma model possesses a similar
property: The flowed $N$-vector field and its composite operators are UV finite
without multiplicative wave function renormalization. Our proof in all orders
of perturbation theory uses a $(2+1)$-dimensional field theoretical
representation of the gradient flow, which possesses local gauge invariance
without gauge field. As application of the UV finiteness of the gradient flow,
we construct the energy--momentum tensor in the lattice formulation of the
$O(N)$ non-linear sigma model that automatically restores the correct
normalization and the conservation law in the continuum limit.
\end{abstract}
\subjectindex{B31, B32, B34, B38}
\maketitle

\section{Introduction and summary}
\label{sec:1}
The Yang--Mills gradient flow or the Wilson flow~\cite{Luscher:2010iy} has
attracted much attention in recent years in the context of lattice gauge
theory. Its known applications include, scale
setting~\cite{Luscher:2010iy,Borsanyi:2012zs}, definition of the topological
charge~\cite{Luscher:2010iy,Noaki:2014ura}, definition of non-perturbative
gauge coupling~\cite{Fodor:2012td,Fritzsch:2013je}, chiral
condensation~\cite{Luscher:2013cpa}, improvement of step
scaling~\cite{Luscher:2014kea}, etc. Even its application to supersymmetric
theory~\cite{Kikuchi:2014rla} and to the operator product
expansion~\cite{Monahan:2014tea} is considered.
Reference~\cite{Luscher:2013vga} is a review of this notion and further related
works can be found in a review~\cite{Aoki:2013ldr} and in a most recent paper
on the non-perturbative beta function~\cite{Hasenfratz:2014rna}.

A crucial property of the Yang--Mills gradient flow, underlying the above
applications, is its ``ultraviolet (UV)
finiteness''~\cite{Luscher:2010iy,Luscher:2011bx}. The gradient flow is a
one-parameter (called the flow-time) evolution of the gauge field, according
to a ``heat diffusion equation'' (called the flow equation). A remarkable fact
that can be rigorously proven~\cite{Luscher:2011bx} in all orders of
perturbation theory is that any correlation function of the evolved (or flowed)
gauge field becomes UV finite \emph{without\/} the wave function
renormalization, as long as the parameters of the theory are properly
renormalized. Moreover, \emph{any\/} local product of the flowed gauge field
remains UV finite without further (multiplicative as well as subtractive)
renormalization. This remarkable property of the gradient flow facilitates, in
particular, the construction of renormalized composite operators of the gauge
field. That is, any simple product of the flowed (bare) gauge field as it
stands is a renormalized (i.e., UV-finite) quantity.

In Ref.~\cite{Suzuki:2013gza}, as possible application of the gradient flow,
one of us (H.S.) considered the construction of the energy--momentum tensor in
lattice gauge theory. This application of the gradient flow to the
energy--momentum tensor was further developed from a somewhat different
perspective in~Ref.~\cite{DelDebbio:2013zaa}. The construction was then
generalized to gauge theories including the fermion
field~\cite{Makino:2014taa}. The genuine energy--momentum tensor cannot be
defined on the lattice because the lattice structure breaks the translational
invariance explicitly. Even the construction of a lattice operator that reduces
to the correctly normalized conserved energy--momentum tensor in the continuum
limit is quite non-trivial as investigated
in~Refs.~\cite{Caracciolo:1988hc,Caracciolo:1989pt}.
Reference~\cite{Fujikawa:1983az} is a pioneering work on this issue.

The basic idea of~Refs.~\cite{Suzuki:2013gza,Makino:2014taa}, which uses the UV
finiteness of the gradient flow in an essential way, is recapitulated
in~Sect.~\ref{sec:6} of the present paper. The aim
of~Refs.~\cite{Suzuki:2013gza,Makino:2014taa} is to construct a lattice
operator that automatically reduces to the correctly normalized conserved
energy--momentum tensor in the continuum limit. Theoretically, there is only
little room for doubt on the reasoning
in~Refs.~\cite{Suzuki:2013gza,Makino:2014taa}. Practically, however, it is not
a priori clear whether presently available lattice parameters are sufficient to
extract physical information by using the construction. On this issue, the
promising result in~Ref.~\cite{Asakawa:2013laa} for thermodynamical quantities
in quenched QCD is quite encouraging. Still, it is indispensable to numerically
demonstrate the conservation law of the energy--momentum tensor by using
lattice Monte Carlo simulations.

Under these situations, it seems useful to study a simpler system that would
allow a similar construction of the lattice energy--momentum tensor using the
gradient flow. One of the basic assumptions
in~Refs.~\cite{Suzuki:2013gza,Makino:2014taa} is that the theory is
asymptotically free. Not so many field theories exhibit asymptotic freedom,
however. This was our original motivation for the present study on the gradient
flow in the 2D $O(N)$ non-linear sigma
model~\cite{Polyakov:1975rr,Migdal:1975zf,Brezin:1975sq}. It is well
known~\cite{Peskin:1995ev} that the physics of this systems possesses many
similarities with the 4D non-Abelian gauge theory. These include asymptotic
freedom, dynamical generation of the mass gap, and, for~$N=3$, the topological
term and associated $\theta$-parameter. See also
Refs.~\cite{Bogli:2011aa,Nogradi:2012dj}. This system is also advantageous from
a computational perspective (and thus from our original motivation), because
there exists a very efficient cluster simulation
algorithm~\cite{Wolff:1988uh,Wolff:1989hv}. The state of the art in
non-perturbative lattice study of the 2D $O(N)$ non-linear sigma model can be
found in~Ref.~\cite{Balog:2012db}.

In the present paper, we will show that there exists another surprising
similarity between the 2D $O(N)$ non-linear sigma model and the 4D gauge
theory: Any correlation function of the flowed $N$-vector field in the former
becomes UV finite \emph{without\/} the wave function renormalization, as long
as the parameters of the theory are renormalized. This UV finiteness also
persists for \emph{any\/} local product of the flowed $N$-vector field. This
similarity is surprising, because the UV finiteness of the flowed gauge field
is a non-trivial consequence~\cite{Luscher:2011bx} of the gauge BRS symmetry
that acts \emph{non-linearly\/} on the gauge field. In fact, matter fields such
as the fermion field transform linearly under the gauge BRS symmetry and they
\emph{do\/} require wave function renormalisation even after the
flow~\cite{Luscher:2013cpa}. In the 2D $O(N)$ non-linear sigma model, however,
it is not clear at first glance what plays the same role as this gauge BRS
symmetry in the 4D gauge theory. Our proof clarifies this point. On the other
hand, happily, because of the UV finiteness of the gradient flow in the 2D
$O(N)$ non-linear sigma model, we can repeat the construction of the lattice
energy--momentum tensor in~Refs.~\cite{Suzuki:2013gza,Makino:2014taa}.

The following describes the organization of the present paper and gives a
summary of the contents of each section.

In Sect.~\ref{sec:2}, we introduce the flow equation in the 2D $O(N)$
non-linear sigma model. If one considers the application in lattice numerical
simulations, this is the equation that should be solved numerically in
conjunction with the conventional Monte Carlo simulations. We then formulate
the perturbative expansion for the system defined by the combination of the 2D
$O(N)$ non-linear sigma model and the flow equation (the flowed system).

In Sect.~\ref{sec:3}, on the basis of the perturbative expansion developed
in~Sect.~\ref{sec:2}, we explicitly compute the two-point function of the
flowed bare $N$-vector field to the one-loop order. This explicit calculation
shows that the two-point function is made UV finite by the conventional
parameter renormalization in the non-linear sigma model~\cite{Brezin:1976ap},
but without the wave function renormalization. We carry out the computation in
dimensional regularization and in lattice regularization and arrive at the same
conclusion. Although this computation is only in the one-loop level, it
strongly indicates that the gradient flow in the non-linear sigma model has a
similar UV property as the gauge theory.

As the proof for the 4D gauge theory in~Ref.~\cite{Luscher:2011bx} and the
renormalizability proof in the stochastic
quantization~\cite{ZinnJustin:1986eq,ZinnJustin:1987ux}, our proof in all
orders of perturbation theory uses a local field theory with one spacetime
dimension higher: We use a $(2+1)$-dimensional field theoretical representation
of the flowed system. In~Sect.~\ref{sec:4}, we define this $(2+1)$-dimensional
local field theory. Then we show that the system defined through the flow
equation in~Sect.~\ref{sec:2} and the $(2+1)$-dimensional field theory have
equivalent perturbative expansions. It is easy to see the rough equivalence.
However, a closer look reveals that there are discrepancies between the two
systems; the measure term in the former is missing in the latter, while the
former does not have the flow-line loop diagrams of the latter. Presumably, the
step to show that these two apparently different elements are actually
equivalent (Sect.~\ref{sec:4.4}) is the hardest part in our argument. We will
find that, to address this very subtle problem in a convincing manner, it is
necessary to first discretize the flow-time derivative and then take the
continuum limit for this discretization; this necessity of discretization is
also counterintuitive.

Once having obtained a local field theory that is (perturbatively) equivalent
to the flowed system, a possible way to proceed is to write down a
Ward--Takahashi relation or a Zinn-Justin equation~\cite{ZinnJustin:1974mc}
(see, e.g., Ref.~\cite{Taylor:1976ru}) for the 1PI generating
functional,\footnote{In this aspect, our approach is more conventional than the
approach in~Ref.~\cite{Luscher:2011bx}.} which restricts the possible form of
counterterms, on the basis of a certain symmetry in the $(2+1)$-dimensional
system. This is the content of~Sect.~\ref{sec:5}. Here, we encounter another
surprise: The $(2+1)$-dimensional field theory possesses \emph{local\/} gauge
symmetries, although it does not contain any gauge field. Note that the unique
internal symmetry in the original 2D $O(N)$ non-linear sigma model is the
global $O(N)$ symmetry. Because of these gauge symmetries, we have to fix the
gauge. Even under the gauge fixing, there still remains a residual symmetry
that acts \emph{non-linearly\/} on various fields. We will find that the
Zinn-Justin equation associated with this non-linear symmetry does the job.
Then, by listing possible counterterms (by borrowing the information obtained
in~Sect.~\ref{sec:4.4}) and examining the restriction implied by the
Zinn-Justin equation, we finally show that the only counterterms required are
those of the original 2D $O(N)$ non-linear sigma model. In particular, the
flowed $N$-vector field (and its composite operators) is not renormalized. This
completes our proof for the UV finiteness of the gradient flow.

In~Sect.~\ref{sec:6}, on the basis of the UV finiteness established
in~Sect.~\ref{sec:5}, we construct the energy--momentum tensor in a lattice
formulation of the non-linear sigma model, following the line of reasoning
of~Refs.~\cite{Suzuki:2013gza,Makino:2014taa}.

In summary, we have found another example in which the gradient flow exhibits a
remarkable UV finiteness: in the 2D $O(N)$ non-linear sigma model, any
correlation function of the flowed $N$-vector field and its composite operators
is UV finite without multiplicative (as well as subtractive) renormalization.
Our proof in the present paper also clarifies subtle but very interesting
technical issues arising in the theoretical analysis of the gradient flow, such
as the necessity of the discretization of the flow-time derivative and the
emergence of gauge and/or non-linear symmetries in the corresponding local
field theory with one dimension higher. The knowledge obtained here will be
useful in considering the application of the gradient flow to a wider range of
systems.

Also, going back to our original motivation, we hope to numerically test the
idea of~Refs.~\cite{Suzuki:2013gza,Makino:2014taa} by using the
energy--momentum tensor constructed in~Sect.~\ref{sec:6} in the near future.

\section{Gradient flow in the 2D $O(N)$ non-linear sigma model}
\label{sec:2}
\subsection{2D $O(N)$ non-linear sigma model and the flow equation}
The 2D $O(N)$ non-linear sigma model is a field theory of an $N$
component vector with the unit length. Its partition function is given
by\footnote{Throughout the present paper, the symbol~$\mathcal{D}$ is
used for the functional integral over functions on the $D$-dimensional
spacetime.}
\begin{equation}
   \mathcal{Z}_{O(N)}
   =\int\left[\prod_{i=1}^N\mathcal{D}n^i\right]
   \left[\prod_x\delta(n(x)^2-1)\right]
   \exp\left[
   -\frac{1}{2g_0^2}\int\mathrm{d}^Dx\,\sum_{i=1}^N
   \partial_\mu n^i(x)\partial_\mu n^i(x)
   \right],
\label{eq:(2.1)}
\end{equation}
where $n(x)^2\equiv\sum_{i=1}^Nn^i(x)n^i(x)$ and~$g_0$ is the bare coupling
constant. Although the spacetime dimension~$D$ is~$2$ for our target theory,
expressions for generic~$D$ are useful because we will extensively use
dimensional regularization in what follows.

In the present paper, as an analogue of the Yang--Mills gradient
flow~\cite{Luscher:2010iy}, we consider the following $t$-evolution of the
$N$-vector field (the flow equation):
\begin{equation}
   \partial_t n^i(t,x)=P^{ij}(t,x)\partial_\mu\partial_\mu n^j(t,x),
\label{eq:(2.2)}
\end{equation}
where the initial value is given by the $N$-vector field in the $O(N)$
non-linear sigma model,
\begin{equation}
   n^i(t=0,x)=n^i(x),
\label{eq:(2.3)}
\end{equation}
which is subject to the functional integral~\eqref{eq:(2.1)}. The projection
operator~$P^{ij}(t,x)$ in the right-hand side of the flow
equation~\eqref{eq:(2.2)} is defined by
\begin{equation}
   P^{ij}(t,x)\equiv\delta^{ij}-n^i(t,x)n^j(t,x)
\label{eq:(2.4)}
\end{equation}
(in~Eq.~\eqref{eq:(2.2)} and in what follows, the sum over the repeated index
is understood). The projection operator is introduced so that the flow is
consistent with the constraint $n(t,x)^2=1$, where
$n(t,x)^2\equiv\sum_{i=1}^Nn^i(t,x)n^i(t,x)$, i.e., $\partial_tn(t,x)^2=0$. The
latter would be a natural requirement for the flow equation for the $O(N)$
non-linear sigma model. In fact, a flow equation identical
to~Eq.~\eqref{eq:(2.2)} has also been advocated in~Appendix~B
of~Ref.~\cite{Kikuchi:2014rla} from the perspective of the symmetry of the
present system.\footnote{It is legitimate to call Eq.~\eqref{eq:(2.2)} the
``gradient'' flow, because the right-hand side of Eq.~\eqref{eq:(2.2)} can also
be obtained as the equation of motion (i.e., the gradient in the functional
space) in the system~\eqref{eq:(2.1)}.}

\subsection{Perturbative expansion}
As usual, for the perturbative treatment of the $O(N)$ non-linear sigma model,
we parametrize the constraint~$n(x)^2=1$ in~Eq.~\eqref{eq:(2.1)} in terms of
$N-1$~independent components (the $\pi$-field) as
\begin{align}
   n^k(x)&=\pi^k(x),\qquad\text{for $k=1$, \dots, $N-1$},
\label{eq:(2.5)}
\\
   n^N(x)&=\sqrt{1-\pi(x)^2},
   \qquad\pi(x)^2\equiv\sum_{k=1}^{N-1}\pi^k(x)\pi^k(x),
\label{eq:(2.6)}
\end{align}
and then expand expressions regarding $\pi(x)$ as a small fluctuation. In this
perturbative treatment, the partition function becomes
\begin{align}
   \mathcal{Z}_{O(N)}
   &=\int\left[\prod_{k=1}^{N-1}\mathcal{D}\pi^k\right]
   \left[\prod_x\frac{1}{\sqrt{1-\pi(x)^2}}\right]
\notag\\
   &\qquad{}
   \times\exp\left(
   -\frac{1}{2g_0^2}\int\mathrm{d}^Dx\,
   \left\{
   \left[\partial_\mu\pi(x)\right]^2
   +\left[\partial_\mu\sqrt{1-\pi(x)^2}\right]^2
   \right\}
   \right).
\label{eq:(2.7)}
\end{align}
The above arbitrary choice of the perturbative branch, Eq.~\eqref{eq:(2.6)}
with small~$\pi(x)$, however, induces infrared (IR) divergences in the
perturbative expansion of $O(N)$ non-invariant quantities~\cite{David:1982qv}.
To regularize the IR divergences, we introduce the mass term
\begin{align}
   S_{\text{mass}}&=-\frac{m_0^2}{g_0^2}
   \int\mathrm{d}^Dx\,\left[n^N(x)-1\right]
\notag\\
   &=\frac{m_0^2}{g_0^2}\int\mathrm{d}^Dx\,
   \left\{\frac{1}{2}\pi(x)^2+\frac{1}{8}\left[\pi(x)^2\right]^2
   +\dotsb\right\},
\label{eq:(2.8)}
\end{align}
and take the massless limit~$m_0\to0$ in the very end of the calculation. With
this mass term, the particular perturbative branch~\eqref{eq:(2.6)} is favored
for a weak coupling.

\begin{figure}
\begin{center}
\includegraphics[width=2cm,clip]{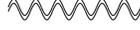}
\caption{A double wavy line represents the heat kernel~\eqref{eq:(2.13)}.}
\label{fig:1}
\end{center}
\end{figure}

Also for the flowed field~$n^i(t,x)$, since $n(t,x)^2=1$ holds along the flow
evolution, we set
\begin{align}
   n^k(t,x)&=\pi^k(t,x),\qquad\text{for $k=1$, \dots, $N-1$},
\label{eq:(2.9)}
\\
   n^N(t,x)&=\sqrt{1-\pi(t,x)^2},
   \qquad\pi(t,x)^2\equiv\sum_{k=1}^{N-1}\pi^k(t,x)\pi^k(t,x).
\label{eq:(2.10)}
\end{align}
Then the perturbative expansion of the flow equation~\eqref{eq:(2.2)} is
obtained from the integral representation,
\begin{equation}
   \pi^k(t,x)
   =\int\mathrm{d}^Dy\,\left[
   K_t(x-y)\pi^k(y)+\int_0^t\mathrm{d}s\,K_{t-s}(x-y)R^k(s,y)\right],
\label{eq:(2.11)}
\end{equation}
where $K_t(x)$ is the heat kernel,\footnote{Throughout the present paper, we
use the abbreviation
\begin{equation}
   \int_p\equiv\int\frac{\mathrm{d}^Dp}{(2\pi)^D}.
\label{eq:(2.12)}
\end{equation}
}
\begin{equation}
   K_t(x)=\int_p\mathrm{e}^{ipx}\mathrm{e}^{-tp^2},
\label{eq:(2.13)}
\end{equation}
and
\begin{equation}
   R^k(t,x)\equiv
   -\pi^k(t,x)\left[
   \pi^l(t,x)\partial_\mu\partial_\mu\pi^l(t,x)
   +\sqrt{1-\pi(t,x)^2}\partial_\mu\partial_\mu\sqrt{1-\pi(t,x)^2}
   \right].
\label{eq:(2.14)}
\end{equation}
Noting that $(\partial_t-\partial_\mu\partial_\mu)K_t(x)=0$
and~$K_{t=0}(x)=\delta^D(x)$, we see that Eq.~\eqref{eq:(2.11)} solves
Eq.~\eqref{eq:(2.2)} with the initial condition~\eqref{eq:(2.3)}. By
iteratively solving Eq.~\eqref{eq:(2.11)} in terms of the initial
value~$\pi^k(y)$, therefore, we have a perturbative solution of the flow
equation. This expansion can be represented diagrammatically (the flow Feynman
diagram~\cite{Luscher:2011bx}) and, throughout this paper, we represent the
heat kernel~\eqref{eq:(2.13)} by a double wavy line in~Fig.~\ref{fig:1}. This
line is also called the ``flow-line propagator'' or simply the ``flow line''.

\begin{figure}
\begin{center}
\includegraphics[width=2cm,clip]{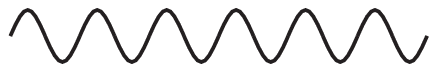}
\caption{A single wavy line represents the free propagator~\eqref{eq:(2.15)}.}
\label{fig:2}
\end{center}
\end{figure}


On the other hand, the combination~$R^k$ in~Eq.~\eqref{eq:(2.14)} represents
the effect of non-linear terms in the flow equation and, in what follows, this
interaction will be denoted by an open circle (the flow vertex); see
Fig.~\ref{fig:4} for an example.

The initial value of the flow, $\pi^k(y)$ in~Eq.~\eqref{eq:(2.11)}, is a
quantum field subject to the functional integral~\eqref{eq:(2.7)}.
From~Eq.~\eqref{eq:(2.11)} and Eq.~\eqref{eq:(2.7)} (with the mass
term~\eqref{eq:(2.8)}), one then sees that the quantum free propagator of the
flowed field is given by
\begin{equation}
   \left\langle\pi^k(t,x)\pi^l(s,y)\right\rangle_0
   =g_0^2\delta^{kl}\int_p\mathrm{e}^{ip(x-y)}
   \frac{\mathrm{e}^{-(t+s)p^2}}{p^2+m_0^2}.
\label{eq:(2.15)}
\end{equation}
Note that in this propagator, the flow times at the end points appear in the
sum (not the difference). Throughout this paper, this free propagator will be
denoted by a single wavy line (Fig.~\ref{fig:2}).

Finally, the functional integral~\eqref{eq:(2.7)} generates interaction
vertices among the $\pi^k(x)$. The interaction vertices in the action integral
will be denoted by a filled circle (see Fig.~\ref{fig:3} for an example). On
the other hand, the interaction vertices arising from the functional measure
in~Eq.~\eqref{eq:(2.7)}, the ``measure term'',
\begin{equation}
   \prod_x
   \frac{1}{\sqrt{1-\pi(x)^2}}
   =\exp\left\{-\frac{1}{2}\delta^D(0)\int\mathrm{d}^Dx\,
   \ln\left[1-\pi(x)^2\right]\right\},
\label{eq:(2.16)}
\end{equation}
will be represented by a cross as in~Fig.~\ref{fig:5}.

\section{One-loop calculation of correlation functions of the flowed field}
\label{sec:3}
An explicit one-loop calculation of the correlation functions of the flowed
field is quite instructive, because it shows a remarkable UV property of the
gradient flow. As the UV regularization, we first adopt dimensional
regularization, setting
\begin{equation}
   D=2-\epsilon.
\label{eq:(3.1)}
\end{equation}

Let us compute the two-point function of the flowed $\pi$-field. The
lowest-order (tree-level) two-point function is given by the free
propagator~\eqref{eq:(2.15)} in~Fig.~\ref{fig:2}.

\begin{figure}
\begin{center}
\includegraphics[width=3cm,clip]{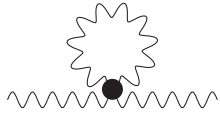}
\caption{Diagram 01: A one-loop diagram that gives rise to the
contribution~\eqref{eq:(3.2)} to the two-point function.}
\label{fig:3}
\end{center}
\end{figure}

\begin{figure}
\begin{center}
\includegraphics[width=3cm,clip]{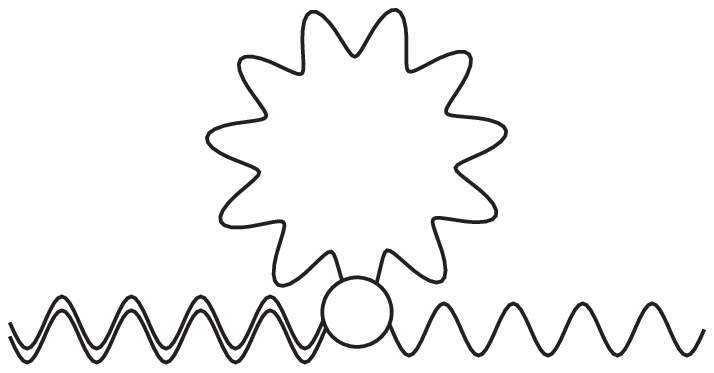}
\caption{Diagram 02: A one-loop diagram that gives rise to the
contribution~\eqref{eq:(3.3)} to the two-point function.}
\label{fig:4}
\end{center}
\end{figure}

\begin{figure}
\begin{center}
\includegraphics[width=3cm,clip]{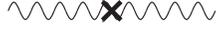}
\caption{Contribution of the measure term to the two-point function. With
lattice regularization, this gives rise to~Eq.~\eqref{eq:(3.10)}.}
\label{fig:5}
\end{center}
\end{figure}


In the one-loop level, diagram~01 in~Fig.~\ref{fig:3}, which contains the
interaction vertex in the original non-linear sigma model only, gives
\begin{align}
   &\left\langle\pi^k(t,x)\pi^l(s,y)\right\rangle
\notag\\
   &=\frac{g_0^2}{4\pi}
   \left[-\frac{2}{\epsilon}
   +\ln\left(\frac{\mathrm{e}^{\gamma_\mathrm{E}}m_0^2}{4\pi}
   \right)\right]
   g_0^2\delta^{kl}\int_p\mathrm{e}^{ip(x-y)}\,
   \frac{\mathrm{e}^{-(t+s)p^2}}{p^2+m_0^2}
\notag\\
   &\qquad{}
   +\frac{g_0^2}{4\pi}\frac{N-3}{2}
   \left[-\frac{2}{\epsilon}
   +\ln\left(\frac{\mathrm{e}^{\gamma_\mathrm{E}}m_0^2}{4\pi}
   \right)\right]
   g_0^2\delta^{kl}\int_p\mathrm{e}^{ip(x-y)}\mathrm{e}^{-(t+s)p^2}\,
   \frac{m_0^2}{(p^2+m_0^2)^2},
\label{eq:(3.2)}
\end{align}
where $\gamma_\mathrm{E}$ is Euler's constant.

On the other hand, the contribution of another one-loop diagram, diagram~02
in~Fig.~\ref{fig:4}, that contains the flow vertex is
\begin{align}
   &\left\langle\pi^k(t,x)\pi^l(s,y)\right\rangle
\notag\\
   &=\frac{g_0^2}{4\pi}(N-1)
   \left[\frac{2}{\epsilon}
   +\frac{1}{2}\ln(8\pi t)+\frac{1}{2}\ln(8\pi s)
   +m_0^2t\ln(2\mathrm{e}^{\gamma_{\mathrm{E}}-1}m_0^2t)
   +m_0^2s\ln(2\mathrm{e}^{\gamma_{\mathrm{E}}-1}m_0^2s)\right]
\notag\\
   &\qquad{}
   \times g_0^2\delta^{kl}\int_p\mathrm{e}^{ip(x-y)}\,
   \frac{\mathrm{e}^{-(t+s)p^2}}{p^2+m_0^2}.
\label{eq:(3.3)}
\end{align}

We note that the measure term in~Eq.~\eqref{eq:(2.16)} vanishes identically in
dimensional regularization with which $\delta^D(0)\equiv0$. Thus, in total, we
have
\begin{align}
   &\left\langle\pi^k(t,x)\pi^l(s,y)\right\rangle
\notag\\
   &=\biggl\{1+\frac{g_0^2}{4\pi}
   \biggl[(N-2)\frac{2}{\epsilon}
   +\ln\left(\frac{\mathrm{e}^{\gamma_\mathrm{E}}m_0^2}{4\pi}\right)
   +\frac{1}{2}(N-1)\ln(8\pi t)+\frac{1}{2}(N-1)\ln(8\pi s)
\notag\\
   &\qquad\qquad\qquad\qquad\qquad{}
   +(N-1)m_0^2t\ln(2\mathrm{e}^{\gamma_{\mathrm{E}}-1}m_0^2t)
   +(N-1)m_0^2s\ln(2\mathrm{e}^{\gamma_{\mathrm{E}}-1}m_0^2s)
   \biggr]\biggr\}
\notag\\
   &\qquad\qquad\qquad\qquad\qquad\qquad{}\times
   g_0^2\delta^{kl}\int_p\mathrm{e}^{ip(x-y)}\,
   \frac{\mathrm{e}^{-(t+s)p^2}}{p^2+m_0^2}
\notag\\
   &\qquad{}
   +\frac{g_0^2}{4\pi}\frac{N-3}{2}
   \left[-\frac{2}{\epsilon}
   +\ln\left(\frac{\mathrm{e}^{\gamma_\mathrm{E}}m_0^2}{4\pi}
   \right)\right]
   g_0^2\delta^{kl}\int_p\mathrm{e}^{ip(x-y)}\mathrm{e}^{-(t+s)p^2}\,
   \frac{m_0^2}{(p^2+m_0^2)^2}+O(g_0^4).
\label{eq:(3.4)}
\end{align}

Now, the parameter renormalization in the original $O(N)$ non-linear sigma
model~\eqref{eq:(2.7)} with the mass term~\eqref{eq:(2.8)} is known to be (in
the minimal subtraction (MS) scheme)
\begin{equation}
   g_0^2\equiv\mu^\epsilon g^2Z,\qquad
   Z=1-\frac{g^2}{4\pi}2(N-2)\frac{1}{\epsilon}+O(g^4),
\label{eq:(3.5)}
\end{equation}
and
\begin{equation}
   m_0^2=\frac{Z}{Z_3^{1/2}}m^2
   =\left[1-\frac{g^2}{4\pi}(N-3)\frac{1}{\epsilon}+O(g^4)\right]m^2,\qquad
   Z_3=1-\frac{g^2}{4\pi}2(N-1)\frac{1}{\epsilon}+O(g^4),
\label{eq:(3.6)}
\end{equation}
where $Z_3$ is the wave function renormalization factor for the unflowed
$\pi$-field, $\pi^k(x)=Z_3^{1/2}\pi_R^k(x)$.\footnote{In Sect.~\ref{sec:5}, as a
byproduct of our analysis, we will have a proof for these renormalization
rules.} If we make these substitutions in~Eqs.~\eqref{eq:(3.4)}, we obtain the
following completely UV-finite expression:
\begin{align}
   &\left\langle\pi^k(t,x)\pi^l(s,y)\right\rangle
\notag\\
   &=\biggl\{1+\frac{g^2}{4\pi}
   \biggl[
   \ln\left(\frac{\mathrm{e}^{\gamma_\mathrm{E}}m^2}{4\pi\mu^2}\right)
   +\frac{1}{2}(N-1)\ln(8\pi\mu^2t)+\frac{1}{2}(N-1)\ln(8\pi\mu^2s)
\notag\\
   &\qquad\qquad\qquad{}
   +(N-1)m^2t\ln(2\mathrm{e}^{\gamma_{\mathrm{E}}-1}m^2t)
   +(N-1)m^2s\ln(2\mathrm{e}^{\gamma_{\mathrm{E}}-1}m^2s)
   \biggr]\biggr\}
\notag\\
   &\qquad\qquad\qquad\qquad{}\times
   g^2\delta^{kl}\int_p\mathrm{e}^{ip(x-y)}\,
   \frac{\mathrm{e}^{-(t+s)p^2}}{p^2+z_mm^2}+O(g^4),
\label{eq:(3.7)}
\end{align}
where
\begin{equation}
   z_m=1-\frac{g^2}{4\pi}\frac{1}{2}(N-3)
   \ln\left(\frac{\mathrm{e}^{\gamma_\mathrm{E}}m^2}{4\pi\mu^2}\right).
\label{eq:(3.8)}
\end{equation}
Remarkably, when expressed in terms of renormalized parameters, the two-point
function of the flowed $\pi$-field is UV finite \emph{without multiplicative
wave function renormalization}.\footnote{Kengo Kikuchi and his collaborators
independently observed this UV finiteness (private communication).} This UV
finiteness of the flowed field is similar to that of the 4D gauge field flowed
by the Yang--Mills gradient flow, a property first observed
in~Ref.~\cite{Luscher:2010iy} in lower-order perturbative computations and then
proven in all orders of perturbation theory in~Ref.~\cite{Luscher:2011bx}. The
above result indicates that by a similar mechanism to the 4D gauge theory, the
$N$-vector field flowed to positive flow times is UV finite only with parameter
renormalization.

It is also instructive to repeat the above calculation by using lattice
regularization instead of dimensional regularization. We adopt the prescription
that in~Eq.~\eqref{eq:(2.7)} $\int\mathrm{d}^Dx\to a^2\sum_x$, where $a$
denotes the lattice spacing, and the derivative~$\partial_\mu$ is replaced by
the forward difference operator. The Laplacian in the flow
equation~\eqref{eq:(2.2)} is replaced by
$\partial_\mu\partial_\mu\to\partial_\mu^*\partial_\mu$, where $\partial_\mu$
and~$\partial_\mu^*$ are the forward and backward difference operators,
respectively. Then the contribution of~Fig.~\ref{fig:3} is
\begin{align}
   &\left\langle\pi^k(t,x)\pi^l(s,y)\right\rangle
\notag\\
   &=\frac{g_0^2}{4\pi}
   \left[\ln(am_0)^2-5\ln2+\pi\right]
   g_0^2\delta^{kl}\int_p\mathrm{e}^{ip(x-y)}\,
   \frac{\mathrm{e}^{-(t+s)p^2}}{p^2+m_0^2}
\notag\\
   &\qquad{}
   +\frac{g_0^2}{4\pi}
   \left\{\frac{N-3}{2}
   \left[\ln(am_0)^2-5\ln2\right]
   -\pi\right\}
   g_0^2\delta^{kl}\int_p\mathrm{e}^{ip(x-y)}\mathrm{e}^{-(t+s)p^2}\,
   \frac{m_0^2}{(p^2+m_0^2)^2}
\notag\\
   &\qquad\qquad{}
   +\frac{g_0^2}{4\pi}\left(-\frac{4\pi}{a^2}\right)
   g_0^2\delta^{kl}\int_p\mathrm{e}^{ip(x-y)}\,
   \frac{\mathrm{e}^{-(t+s)p^2}}{(p^2+m_0^2)^2},
\label{eq:(3.9)}
\end{align}
which is quadratically divergent. The quadratic divergence in the last term is
canceled by the measure term~\eqref{eq:(2.16)} with~$\delta^D(0)\to1/a^2$ for
lattice regularization. In fact, the contribution of the measure term to the
two-point function (Fig.~\ref{fig:5}) is
\begin{equation}
   \left\langle\pi^k(t,x)\pi^l(s,y)\right\rangle
   =\frac{g_0^2}{4\pi}\left(\frac{4\pi}{a^2}\right)
   g_0^2\delta^{kl}\int_p\mathrm{e}^{ip(x-y)}\,
   \frac{\mathrm{e}^{-(t+s)p^2}}{(p^2+m_0^2)^2}.
\label{eq:(3.10)}
\end{equation}

On the other hand, the contribution of~Fig.~\ref{fig:4} is
\begin{align}
   &\left\langle\pi^k(t,x)\pi^l(s,y)\right\rangle
\notag\\
   &=\frac{g_0^2}{4\pi}(N-1)
   \left[-\ln(am_0)^2+5\ln2
   +\frac{1}{2}\ln(2\mathrm{e}^{\gamma_{\mathrm{E}}}m_0^2t)
   +\frac{1}{2}\ln(2\mathrm{e}^{\gamma_{\mathrm{E}}}m_0^2s)\right]
\notag\\
   &\qquad{}
   \times g_0^2\delta^{kl}\int_p\mathrm{e}^{ip(x-y)}\,
   \frac{\mathrm{e}^{-(t+s)p^2}}{p^2+m_0^2}.
\label{eq:(3.11)}
\end{align}

Thus, we have in total
\begin{align}
   &\left\langle\pi^k(t,x)\pi^l(s,y)\right\rangle
\notag\\
   &=\biggl\{1+\frac{g_0^2}{4\pi}
   \biggl[-(N-2)\left[\ln(am_0)^2-5\ln2\right]+\pi
\notag\\
   &\qquad\qquad\qquad{}
   +\frac{1}{2}(N-1)\ln(2\mathrm{e}^{\gamma_{\mathrm{E}}}m_0^2t)
   +\frac{1}{2}(N-1)\ln(2\mathrm{e}^{\gamma_{\mathrm{E}}}m_0^2s)
   \biggr]\biggr\}
   g_0^2\delta^{kl}\int_p\mathrm{e}^{ip(x-y)}\,
   \frac{\mathrm{e}^{-(t+s)p^2}}{p^2+m_0^2}
\notag\\
   &\qquad{}
   +\frac{g_0^2}{4\pi}
   \left\{\frac{N-3}{2}\left[\ln(am_0)^2-5\ln2\right]-\pi\right\}
   g_0^2\delta^{kl}\int_p\mathrm{e}^{ip(x-y)}\mathrm{e}^{-(t+s)p^2}\,
   \frac{m_0^2}{(p^2+m_0^2)^2}+O(g_0^4).
\label{eq:(3.12)}
\end{align}
It is obvious that all UV divergences are removed by the parameter
renormalization~\eqref{eq:(3.5)} and~\eqref{eq:(3.6)} with the replacement
$1/\epsilon\to-\ln a$; again, remarkably, no wave function renormalization is
required.

Although the two-point function~\eqref{eq:(3.7)} is UV finite, it contains IR
divergences (i.e., it diverges for~$m\to0$) because it is not an $O(N)$
invariant ``physical'' quantity~\cite{David:1982qv}. As a simple example of an
IR-finite $O(N)$-invariant observable, we can consider the ``energy density'',
defined by
\begin{equation}
   E(t,x)\equiv\frac{1}{2}\partial_\mu n^i(t,x)\partial_\mu n^i(t,x),
\label{eq:(3.13)}
\end{equation}
which is analogous to the energy density introduced
in~Ref.~\cite{Luscher:2010iy} for the gauge theory.

\begin{figure}
\begin{center}
\includegraphics[width=2cm,clip]{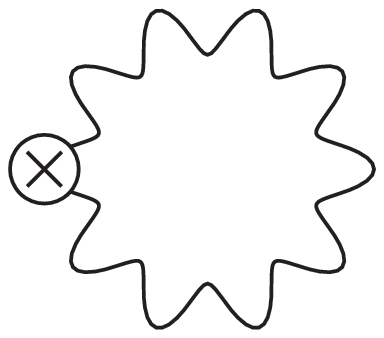}
\caption{Diagram 03: A one-loop diagram that contributes to Eq.~\eqref{eq:(3.15)}.}
\label{fig:6}
\end{center}
\end{figure}

\begin{figure}
\begin{center}
\includegraphics[width=3cm,clip]{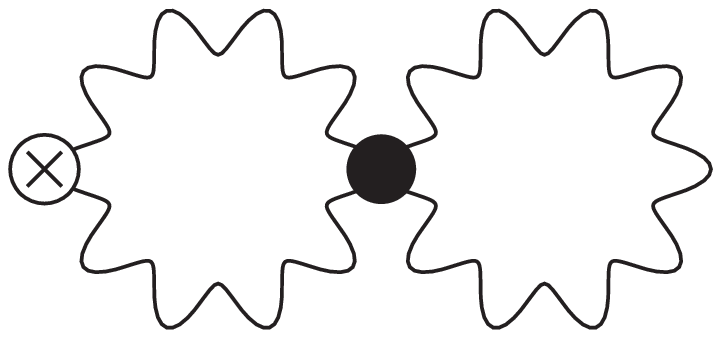}
\caption{Diagram 04: A two-loop diagram that contributes to Eq.~\eqref{eq:(3.15)}.}
\label{fig:7}
\end{center}
\end{figure}

\begin{figure}
\begin{center}
\includegraphics[width=3cm,clip]{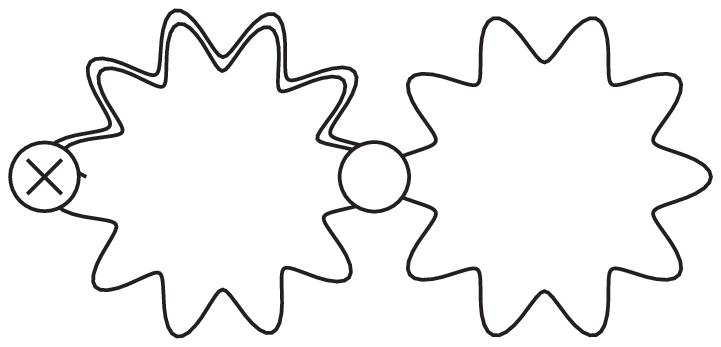}
\caption{Diagram 05: A two-loop diagram that contributes to Eq.~\eqref{eq:(3.15)}.}
\label{fig:8}
\end{center}
\end{figure}

\begin{figure}
\begin{center}
\includegraphics[width=3cm,clip]{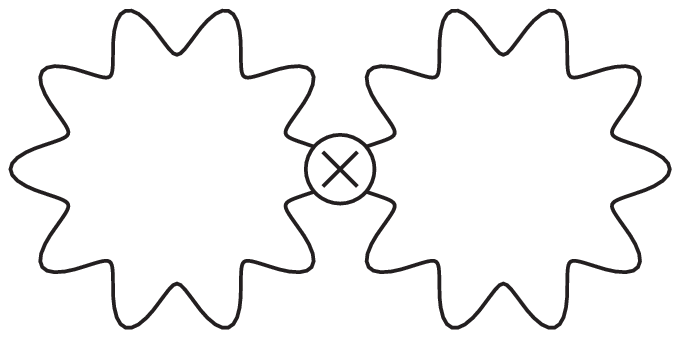}
\caption{Diagram 06: A two-loop diagram that contributes to Eq.~\eqref{eq:(3.15)}.}
\label{fig:9}
\end{center}
\end{figure}


For the vacuum expectation value,
\begin{equation}
   \left\langle E(t,x)\right\rangle
   =\left\langle
   \frac{1}{2}
   \left\{
   \left[\partial_\mu\pi(t,x)\right]^2
   +\left[\partial_\mu\sqrt{1-\pi(t,x)^2}\right]^2
   \right\}
   \right\rangle,
\label{eq:(3.14)}
\end{equation}
there are four flow Feynman diagrams to the next-to-leading order, as depicted
in~Figs.~\ref{fig:6}--\ref{fig:9} (the cross denotes the operator~$E(t,x)$). A
straightforward calculation using dimensional regularization yields
\begin{align}
   \left\langle E(t,x)\right\rangle
   &=\frac{g_0^2}{4\pi}(N-1)\frac{1}{4t}(8\pi t)^{\epsilon/2}
   \left[1+\frac{g_0^2}{4\pi}2(N-2)\frac{1}{\epsilon}(8\pi t)^{\epsilon/2}
   +O(g_0^4)\right]
\notag\\
   &=\frac{g^2}{4\pi}(N-1)\frac{1}{4t}
   \left[1+\frac{g^2}{4\pi}(N-2)\ln(8\pi\mu^2 t)+O(g^4)\right].
\label{eq:(3.15)}
\end{align}
This is IR finite as expected and UV finite in terms of the renormalized
coupling constant, again indicating the UV finiteness of the flowed field. If
this UV finiteness persists to all orders (we will prove this in a later
section), the result~\eqref{eq:(3.15)} shows that the combination
$t\langle E(t,x)\rangle$ provides a possible non-perturbative definition of a
renormalized coupling as the gradient flow scheme in the 4D gauge theory (see,
e.g., Refs.~\cite{Fodor:2012td,Fritzsch:2013je}). That is, we can set
\begin{equation}
   g_R^2(1/\sqrt{8t})
   \equiv\frac{16\pi}{N-1}t\left\langle E(t,x)\right\rangle
   =g_0^2+\dotsb.
\end{equation}
Then it must be interesting to investigate the running of this non-perturbative
coupling in numerical lattice simulations, in view of the expected conformal
and walking behaviors of the $O(3)$ non-linear sigma model with non-zero
$\theta$-parameters~\cite{Nogradi:2012dj}.

\section{$(D+1)$-dimensional field theoretical representation of the gradient flow}
\label{sec:4}
In the next section, we reveal the renormalization structure of the flowed
system defined in~Sect.~\ref{sec:2}. We prove in particular that the flowed
$N$-vector field does not require the wave function renormalization. Our
strategy is identical to the case of the 4D gauge theory~\cite{Luscher:2011bx};
we seek a $(D+1)$-dimensional local field theory that reproduces the flow
Feynman rules in the preceding sections and use this to show the
renormalizability. We neglect the IR-regulating mass term~\eqref{eq:(2.8)} in
this section, because it complicates the argument destroying the $O(N)$
symmetry. We will consider the effect of the mass term at the very end of the
next section.

\subsection{Partition function}
As in~Refs.~\cite{Luscher:2011bx,Luscher:2013cpa}, we consider a
$(D+1)$-dimensional ($D=2$ for our target theory) field theory defined in the
half space, $(t,x)\in[0,\infty)\times\mathbb{R}^D$, that (at least
perturbatively) is equivalent to the gradient flow in the 2D $O(N)$ non-linear
sigma model. We will find that, to resolve subtleties associated with the
measure term and the flow-line loop (see~Sect.~\ref{sec:4.4}), it is necessary
to specify a prescription for the flow-time derivative. We will use the forward
difference prescription (with the discretization length~$\epsilon$) for
this.\footnote{Our renormalization proof uses a $(D+1)$-dimensional system that
assumes a particular forward difference for the flow-time derivative. We do not
mean, however, that the time evolution in the gradient flow must be defined by
the forward time difference; any sound discretization of the flow-time
derivative can be used to implement the flow equation~\eqref{eq:(2.2)} in
numerical simulations. The $(D+1)$-dimensional system below is merely an
intermediate tool for the renormalization proof and, in our present context, is
not an object to be simulated.} The regularization for the $D$-dimensional
``spacetime'' direction is, on the other hand, arbitrary and we may assume, for
instance, dimensional regularization or lattice regularization.

The partition function of the $(D+1)$-dimensional field theory that we consider
is defined by
\begin{align}
   \mathcal{Z}&\equiv\int
   \left[\prod_{i=1}^N\mathcal{D}\xi^i\right]
   \left[\prod_{i=1}^N\mathcal{D}n^i\right]
   \left[\prod_x\delta(n(x)^2-1)\right]
\notag\\
   &\qquad{}
   \times
   \left[\prod_{t=0}^\infty\prod_{i=1}^N\mathcal{D}\lambda^i(t)\right]   
   \left[\prod_{t=0}^\infty\prod_{i=1}^N\mathcal{D}n^i(t)\right]
   \left[\prod_{t=0}^\infty\prod_x
   \delta(n(t,x)^2-1)\sqrt{1-n_\bot(t+\epsilon,x)^2}\right]\,
   \mathrm{e}^{-S},
\label{eq:(4.1)}
\end{align}
where $t=0$, $\epsilon$, $2\epsilon$, \dots, and
\begin{align}
   S&\equiv
   \frac{1}{2g_0^2}\int\mathrm{d}^Dx\,\partial_\mu n^i(x)\partial_\mu n^i(x)
\notag\\
   &\qquad{}
   -i\epsilon\sum_{t=0}^\infty
   \int\mathrm{d}^Dx\,\lambda^i(t,x)
   P^{ij}(t,x)\left\{
   \frac{1}{\epsilon}\left[n^j(t+\epsilon,x)-n^j(t,x)\right]
   -\partial_\mu\partial_\mu n^j(t,x)
   \right\}
\notag\\
   &\qquad\qquad{}
   -i\int\mathrm{d}^Dx\,\xi^i(x)\left[n^i(0,x)-n^i(x)\right].
\label{eq:(4.2)}
\end{align}
In these expressions, $n^i(x)$ corresponds to the $N$-vector field in the
$D$-dimensional $O(N)$ non-linear sigma model~\eqref{eq:(2.1)} and $n^i(t,x)$
corresponds to the $N$-vector field evolved by the flow
equation~\eqref{eq:(2.2)}. The basic idea is that the functional integral over
the Lagrange multiplier~$\lambda^i(t,x)$ imposes the flow
equation~\eqref{eq:(2.2)} with the discretized flow time. Note that the
left-hand side of~Eq.~\eqref{eq:(2.2)} can equivalently be written
as~$P^{ij}(t,x)\partial_tn^j(t,x)$ with the projection operator~$P^{ij}(t,x)$
in~Eq.~\eqref{eq:(2.4)}. The integration over another Lagrange
multiplier~$\xi^i(x)$ in~Eq.~\eqref{eq:(4.1)}, on the other hand, imposes the
initial condition~\eqref{eq:(2.3)}.

In~Eq.~\eqref{eq:(4.1)},
$n_\bot(t+\epsilon,x)^2%
\equiv\sum_{i=1}^Nn_\bot^i(t+\epsilon,x)n_\bot^i(t+\epsilon,x)$, and
\begin{equation}
   n_\bot^i(t+\epsilon,x)\equiv
   \epsilon P^{ij}(t,x)\partial_\mu\partial_\mu n^j(t,x).
\label{eq:(4.3)}
\end{equation}
It can be shown that, with the factor~$\sqrt{1-n_\bot(t+\epsilon,x)^2}$ in the
integration measure, the partition function~$\mathcal{Z}$~\eqref{eq:(4.1)} can
be obtained from the original partition
function~$\mathcal{Z}_{O(N)}$~\eqref{eq:(2.1)} by inserting unity (up to
infinite gauge volume; see below). However, since
$\sqrt{1-n_\bot(t+\epsilon,x)^2}=1+O(\epsilon^2)\to1$ for~$\epsilon\to0$, this
factor can be neglected in the $\epsilon\to0$ limit and we do not explicitly
include this factor in what follows.

\subsection{Symmetries and the gauge fixing}
The above $(D+1)$-dimensional system possesses the following symmetries. One is
the global $O(N)$ symmetry that is inherited from the original $O(N)$
non-linear sigma model:
\begin{align}
   \delta n^i(x)&=\epsilon^{ij}n^j(x),&
   \delta\xi^i(x)&=\epsilon^{ij}\xi^j(x),
\notag\\
   \delta n^i(t,x)&=\epsilon^{ij}n^j(t,x),&
   \delta\lambda^i(t,x)&=\epsilon^{ij}\lambda^j(t,x),
\label{eq:(4.4)}
\end{align}
where $\epsilon^{ji}=-\epsilon^{ij}$ are infinitesimal constant parameters.

Other, somewhat unexpected ones are \emph{local\/} gauge symmetries:
\begin{align}
   \delta n^i(x)&=0,&
   \delta\xi^i(x)&=g(x)\Bar{n}^i(x),
\notag\\
   \delta n^i(t,x)&=0,&
   \delta\lambda^i(t,x)&=h(t,x)n^i(t,x),
\label{eq:(4.5)}
\end{align}
where
\begin{equation}
   \overline{n}^i(x)\equiv\frac{n^i(0,x)+n^i(x)}{2},
\label{eq:(4.6)}
\end{equation}
and $g(x)$ and~$h(t,x)$ are \emph{local\/} parameters that can depend on their
arguments. These local symmetries, which exist even with the discretized
flow-time and $D$-dimensional regularization, follow from the
constraints~$n(x)^2=n(t,x)^2=1$ in the functional integral and the
property~$n^i(t,x)P^{ij}(t,x)=0$. Because of these gauge symmetries, the
partition function~\eqref{eq:(4.1)} itself is infinite. This is not a problem
in our present context, because what we need at this moment is a generating
functional of the perturbative expansion of the flowed system.

To formulate perturbation theory in the above $(D+1)$-dimensional field theory,
we thus have to first fix the gauge symmetries~\eqref{eq:(4.5)}. For this, we
adopt the following gauge fixing conditions,
\begin{equation}
   \xi^N(x)=0,\qquad\lambda^N(t,x)=0,
\label{eq:(4.7)}
\end{equation}
and follow the Faddeev--Popov procedure. Thus we insert unity
\begin{align}
   &\int
   \mathcal{D}g
   \left[\prod_x\delta(\xi^N(x)-g(x)\overline{n}^N(x))
   \left|\overline{n}^N(x)\right|\right]
\notag\\
   &\qquad\qquad{}\times
   \left[\prod_{t=0}^\infty\mathcal{D}h(t)\right]
   \left[\prod_{t=0}^\infty\prod_x\delta(\lambda^N(t,x)-h(t,x)n^N(t,x))
   \left|n^N(t,x)\right|\right]=1
\label{eq:(4.8)}
\end{align}
into the functional integral~\eqref{eq:(4.1)}. Then, using the invariance of
the action and the functional measure under the
transformations~\eqref{eq:(4.5)}, we can factor out the gauge volume
\begin{equation}
   \int\mathcal{D}g\left[\prod_{t=0}^\infty\mathcal{D}h(t)\right]
\label{eq:(4.9)}
\end{equation}
from the partition function~\eqref{eq:(4.1)}.

We further solve the constraints $n(x)^2=n(t,x)^2=1$ in terms of $N-1$
independent components, as~Eqs.~\eqref{eq:(2.5)} and~\eqref{eq:(2.6)}
and~Eqs.~\eqref{eq:(2.9)} and~\eqref{eq:(2.10)}. Then, after the gauge volume
is factored out, the partition function is given by
\begin{align}
   \mathcal{Z}'&=
   \int
   \left[\prod_{k=1}^{N-1}\mathcal{D}\xi^k\right]
   \left[\prod_{k=1}^{N-1}\mathcal{D}\pi^k\right]
\notag\\
   &\qquad{}\times
   \left[\prod_{t=0}^\infty\prod_{k=1}^{N-1}\mathcal{D}\lambda^k(t)\right]
   \left[\prod_{t=0}^\infty\prod_{k=1}^{N-1}\mathcal{D}\pi^k(t)\right]\,
   \prod_x\frac{\overline{\sqrt{1-\pi(x)^2}}}{\sqrt{1-\pi(x)^2}}\,
   \mathrm{e}^{-S},
\label{eq:(4.10)}
\end{align}
where
\begin{equation}
   \overline{\sqrt{1-\pi(x)^2}}
   \equiv\frac{\sqrt{1-\pi(0,x)^2}+\sqrt{1-\pi(x)^2}}{2},
\label{eq:(4.11)}
\end{equation}
and
\begin{align}
   S&=
   \frac{1}{2g_0^2}\int\mathrm{d}^Dx\,\left\{
   \left[\partial_\mu\pi(x)\right]^2
   +\left[\partial_\mu\sqrt{1-\pi(x)^2}\right]^2\right\}
\notag\\
   &\qquad{}-i\epsilon\sum_{t=0}^\infty\int\mathrm{d}^Dx\,\lambda^k(t,x)
   \left(
   \frac{1}{\epsilon}\left[\pi^k(t+\epsilon,x)-\pi^k(t,x)\right]
   -\partial_\mu\partial_\mu\pi^k(t,x)
   -R^k(t,x)\right)
   +\mathcal{E}
\notag\\
   &\qquad\qquad{}-i\int\mathrm{d}^Dx\,\xi^k(x)
   \left[\pi^k(0,x)-\pi^k(x)\right],
\label{eq:(4.12)}
\end{align}
where the combination~$R^k(t,x)$ is defined by~Eq.~\eqref{eq:(2.14)} and
\begin{align}
   \mathcal{E}&\equiv
   i\epsilon\sum_{t=0}^\infty\int\mathrm{d}^Dx\,\lambda^k(t,x)
   \pi^k(t,x)
   \biggl\{
   \pi^l(t,x)\frac{1}{\epsilon}\left[\pi^l(t+\epsilon,x)-\pi^l(t,x)\right]
\notag\\
   &\qquad\qquad\qquad{}+\sqrt{1-\pi(t,x)^2}
   \frac{1}{\epsilon}
   \left[\sqrt{1-\pi(t+\epsilon,x)^2}-\sqrt{1-\pi(t,x)^2}\right]
   \biggr\}.
\label{eq:(4.13)}
\end{align}

\subsection{Feynman rules in the $(D+1)$-dimensional system}
Next we derive the Feynman rules in the above
system~\eqref{eq:(4.10)}--\eqref{eq:(4.13)}. To write down the free
propagator, we introduce the heat kernel with the discretized flow time, by
\begin{equation}
   K_t^\epsilon(x)\equiv\int_p\mathrm{e}^{ipx}(1-\epsilon p^2)^{t/\epsilon},
\label{eq:(4.14)}
\end{equation}
which fulfills
\begin{equation}
   \frac{1}{\epsilon}\left[K_{t+\epsilon}^\epsilon(x)-K_t^\epsilon(x)\right]
   -\partial_\mu\partial_\mu K_t^\epsilon(x)=0,\qquad
   K_0^\epsilon(x)=\delta^D(x).
\label{eq:(4.15)}
\end{equation}
Clearly, $K_t^\epsilon(x)$ reduces to the heat kernel~\eqref{eq:(2.13)} in the
continuum flow-time limit, $K_t^\epsilon(x)\stackrel{\epsilon\to0}{\to}K_t(x)$.
By using this object, we change the integration variables from~$\pi^k(t,x)$
to~$p^k(t,x)$ as~\cite{Luscher:2011bx}
\begin{equation}
   \pi^k(t,x)=\int\mathrm{d}^Dy\,K_t^\epsilon(x-y)\,\pi^k(y)+p^k(t,x).
\label{eq:(4.16)}
\end{equation}
Then the action becomes
\begin{align}
   S&=
   \frac{1}{2g_0^2}\int\mathrm{d}^Dx\,\left\{
   \left[\partial_\mu\pi(x)\right]^2+\dotsb\right\}
\notag\\
   &\qquad{}-i\epsilon\sum_{t=0}^\infty\int\mathrm{d}^Dx\,\lambda^k(t,x)
   \left\{
   \frac{1}{\epsilon}\left[p^k(t+\epsilon,x)-p^k(t,x)\right]
   -\partial_\mu\partial_\mu p^k(t,x)+\dotsb
   \right\}+\dotsb
\notag\\
   &\qquad\qquad{}
   -i\int\mathrm{d}^Dx\,\xi^k(x)p^k(0,x),
\label{eq:(4.17)}
\end{align}
where abbreviated terms are cubic or higher in fields. It is then
straightforward to find free propagators and the result is
\begin{align}
   \left\langle\pi^k(x)\pi^l(y)\right\rangle_0
   &=g_0^2\delta^{kl}\int_p\mathrm{e}^{ip(x-y)}\frac{1}{p^2},
\label{eq:(4.18)}
\\
   \left\langle p^k(t,x)\lambda^l(s,y)\right\rangle_0
   &=i\delta^{kl}\vartheta(t-s)K_{t-s-\epsilon}^\epsilon(x-y),
\label{eq:(4.19)}
\\
   \left\langle p^k(t,x)\xi^l(y)\right\rangle_0
   &=i\delta^{kl}\vartheta(t+\epsilon)K_t^\epsilon(x-y),
\label{eq:(4.20)}
\end{align}
where $\vartheta(t)$ is a ``regularized'' step function,
\begin{equation}
   \vartheta(t)\equiv
   \begin{cases}
   1,&\text{for $t>0$},\\
   0,&\text{for $t=0$},\\
   0,&\text{for $t<0$}.\\
   \end{cases}
\label{eq:(4.21)}
\end{equation}
Note that $\vartheta(0)=0$ (not, e.g., $1/2$). Since other free propagators
among $\pi^k(x)$, $p^k(t,x)$, $\lambda^k(t,x)$, and~$\xi^k(x)$ vanish,
Eqs.~\eqref{eq:(4.18)}--\eqref{eq:(4.20)} in conjunction
with~Eq.~\eqref{eq:(4.16)} show,
\begin{align}
   \left\langle\pi^k(t,x)\pi^l(s,y)\right\rangle_0
   &=g_0^2\delta^{kl}\int_p\mathrm{e}^{ip(x-y)}
   \frac{(1-\epsilon p^2)^{(t+s)/\epsilon}}{p^2},
\label{eq:(4.22)}
\\
   \left\langle\pi^k(t,x)\lambda^l(s,y)\right\rangle_0
   &=i\delta^{kl}\vartheta(t-s)K_{t-s-\epsilon}^\epsilon(x-y),
\label{eq:(4.23)}
\\
   \left\langle\pi^k(t,x)\xi^l(y)\right\rangle_0
   &=i\delta^{kl}\vartheta(t+\epsilon)K_t^\epsilon(x-y).
\label{eq:(4.24)}
\end{align}
In passing, we note
\begin{equation}
   \left\langle\pi^k(t+\epsilon,x)\lambda^l(t,y)\right\rangle_0
   =i\delta^{kl}\delta^D(x-y),\qquad
   \left\langle\pi^k(t,x)\lambda^l(t,y)\right\rangle_0=0,
\label{eq:(4.25)}
\end{equation}
and
\begin{equation}
   \left\langle\pi^k(0,x)\xi^l(y)\right\rangle_0
   =i\delta^{kl}\delta^D(x-y).
\label{eq:(4.26)}
\end{equation}
This completes our derivation of free propagators. In the continuum flow-time
limit~$\epsilon\to0$, the $\pi\pi$-propagator~\eqref{eq:(4.22)} reproduces the
$\pi\pi$-propagator in~Eq.~\eqref{eq:(2.15)} and the
$\pi\lambda$-propagator~\eqref{eq:(4.23)} reproduces the flow-line propagator
$K_{t-s}(x-y)$ in~Eq.~\eqref{eq:(2.11)}; the step function $\vartheta(t-s)$ is
implicitly implied in~Eq.~\eqref{eq:(2.11)} through the retarded time-ordering,
$t>s$.

The interaction terms in the present $(D+1)$-dimensional system are given by
terms in~Eq.~\eqref{eq:(4.12)} being cubic or higher in fields. The first line
of~Eq.~\eqref{eq:(4.12)} of course reproduces the interaction terms in the
action of the $O(N)$ non-linear sigma model, Eq.~\eqref{eq:(2.7)}. On the other
hand, the
term~$i\epsilon\sum_{t=0}^\infty\int\mathrm{d}^Dx\,\lambda^k(t,x)R^k(t,x)$
in the limit~$\epsilon\to0$, combined with the above $\pi\lambda$-propagator,
precisely reproduces the last term of the integral equation~\eqref{eq:(2.11)}
(i.e., the flow vertex).

Thus, we have observed that our present $(D+1)$-dimensional system basically
reproduces the perturbative expansion of the flowed system defined
in~Sect.~\ref{sec:2}; they seem to be basically equivalent. Nevertheless, we
should note that the equivalence appears not quite complete. The measure
term~\eqref{eq:(2.16)} is missing in~Eqs.~\eqref{eq:(4.10)}--\eqref{eq:(4.13)}
(the factor $\prod_x\overline{\sqrt{1-\pi(x)^2}}/\sqrt{1-\pi(x)^2}$ becomes
unity under the integration over~$\xi$ and this is not the measure term).
Although the measure term~\eqref{eq:(2.16)} identically vanishes when one uses
dimensional regularization, it plays an important role in other
regularizations, such as lattice regularization. If the equivalence including
the measure term does not hold, then the renormalizability proof in the next
section, which is based on the present $(D+1)$-dimensional field theory, does
not apply to the gradient flow with, e.g., lattice regularization. Then, the UV
finiteness of the gradient flow with lattice regularization, which we observed
through an explicit calculation in~Sect.~\ref{sec:3}, is not explained by the
proof.

We will find that, rather surprisingly, the measure term is generated from
naively $O(\epsilon)$ terms in the action~\eqref{eq:(4.12)}. The aim of the
next subsection is to clarify this point and to establish the perturbative
equivalence between the above $(D+1)$-dimensional system and the flowed system
in~Sect.~\ref{sec:2}.

\subsection{Equivalence with the perturbative expansion of the gradient flow}
\label{sec:4.4}
We first integrate over the Lagrange multiplier~$\xi^k(x)$ in the partition
function~\eqref{eq:(4.10)}. Then $\pi^k(0,x)$ is identified with~$\pi^k(x)$ and
we have $\prod_x\overline{\sqrt{1-\pi(x)^2}}/\sqrt{1-\pi(x)^2}=1$
in~Eq.~\eqref{eq:(4.10)}.

\begin{figure}
\begin{center}
\includegraphics[width=3cm,clip]{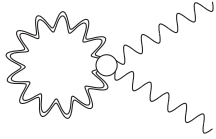}
\caption{An example of the flow-line loop diagram.}
\label{fig:10}
\end{center}
\end{figure}

\begin{figure}
\begin{center}
\includegraphics[width=3cm,clip]{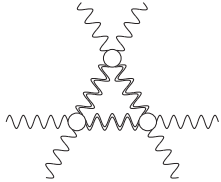}
\caption{Another example of the flow-line loop diagram.}
\label{fig:11}
\end{center}
\end{figure}

Next we note that the perturbative expansion of Eq.~\eqref{eq:(4.10)} generates
loop diagrams consisting solely of the flow-line propagator~\eqref{eq:(4.23)}.
Such ``flow-line loop diagrams'' are depicted in~Figs.~\ref{fig:10}
and~\ref{fig:11}.\footnote{The flow-line loops cannot become higher than
one-loop, because the flow vertex is linear in~$\lambda$.}

It is now very important to recognize that there is no counterpart to the above
flow-line loop diagrams in the perturbative expansion of the flowed system
in~Sect.~\ref{sec:2}. This is a consequence of the retarded nature of the flow
equation and one can confirm this by drawing flow-line diagrams starting
from~Eq.~\eqref{eq:(2.11)}. Thus, there appears some (apparent; see below)
discrepancy between the perturbative expansions of the above two systems.

Let us begin our investigation from the flow-line loop diagram
in~Fig.~\ref{fig:10} which starts and ends at the same flow vertex. First note
that the $\pi\lambda$-propagator~\eqref{eq:(4.23)} vanishes when the flow time
of~$\lambda$ is greater or \emph{equal\/} to the flow time of~$\pi$. Therefore,
in~Eq.~\eqref{eq:(4.12)}, the genuine flow vertex containing the non-linear
term~$R^k(t,x)$ does not contribute to the flow-line loop diagram
in~Fig.~\ref{fig:10}. What contributes is the self-contraction in the
combination~$\mathcal{E}$~\eqref{eq:(4.13)}. The self-contraction
of~$-\mathcal{E}$ yields
\begin{equation}
   \epsilon\sum_{t=0}^\infty\delta^D(0)\int\mathrm{d}^Dx\,
   \frac{1}{\epsilon}
   \left\{\pi^k(t,x)
   \left[\pi^k(t,x)
   -\frac{\sqrt{1-\pi(t,x)^2}}{\sqrt{1-\pi(t+\epsilon,x)^2}}
   \,\pi^k(t+\epsilon,x)\right]\right\}.
\label{eq:(4.27)}
\end{equation}
If we Taylor expand $\pi^k(t+\epsilon,x)$ in this expression with respect
to~$\epsilon$, we find
\begin{equation}
   \int_0^\infty\mathrm{d}t\,\frac{1}{2}
   \delta^D(0)\int\mathrm{d}^Dx\,
   \partial_t\ln\left[1-\pi(t,x)^2\right]+O(\epsilon).
\label{eq:(4.28)}
\end{equation}
Since this is a total derivative, only the boundary field
$\pi^k(t=0,x)=\pi^k(x)$ is contained. Then, remarkably, Eq.~\eqref{eq:(4.28)}
coincides with the measure term~\eqref{eq:(2.16)} for~$\epsilon\to0$.

The above result~\eqref{eq:(4.28)} can be obtained in a somewhat different
manner. We first Taylor expand $-\mathcal{E}$, which yields
\begin{equation}
   i\int_0^\infty\mathrm{d}t\,\int\mathrm{d}^Dx\,
   \left(\frac{1}{2}\lambda^k(t,x)\pi^k(t,x)
   \left\{
   \left[\partial_t\pi^l(t,x)\right]^2
   +\frac{\left[\pi^l(t,x)\partial_t\pi^l(t,x)\right]^2}{1-\pi(t,x)^2}
   \right\}\epsilon+O(\epsilon^2)\right),
\label{eq:(4.29)}
\end{equation}
which is~$O(\epsilon)$. This $O(\epsilon)$ term becomes~$O(1)$ under the
self-contraction, because in the $\epsilon\to0$~limit of the
$\pi\lambda$-propagator~\eqref{eq:(4.23)} behaves as
\begin{equation}
   \left\langle\partial_t\pi^k(t,x)\lambda^l(s,y)\right\rangle_0
   =i\delta^{kl}\delta(t-s)\delta^D(x-y)
   +i\delta^{kl}\theta(t-s)\partial_tK_{t-s}(x-y),
\label{eq:(4.30)}
\end{equation}
and the delta function at the equal flow-time is interpreted
as~$\delta(0)=1/\epsilon$. The self-contraction in~Eq.~\eqref{eq:(4.29)} thus
cancels the factor~$\epsilon$ and leaves the $O(1)$ result,
Eq.~\eqref{eq:(4.28)}.

Next, we see that a flow-line loop diagram that contains a plurality of flow
vertices, such as the diagram in~Fig.~\ref{fig:11}, vanishes as~$\epsilon\to0$.
A little thought shows that \emph{all\/} vertices in such a flow-line loop
diagram must be the vertex arising from~$\mathcal{E}$~\eqref{eq:(4.13)}. This
is again because the $\pi\lambda$-propagator~\eqref{eq:(4.23)} vanishes when
the flow time of~$\lambda$ is greater than or equal to that of~$\pi$. The
vertex is~$O(\epsilon)$ as in~Eq.~\eqref{eq:(4.29)}.

The integration of the flow time of each vertex eliminates one delta function
and finally one is left with an overall integration and~$\delta(0)=1/\epsilon$.
In the present case of a plurality of flow vertices, however, the power 
of~$\epsilon$ coming from the vertices is always greater than or equal to two;
thus the flow-line loop diagram vanishes for~$\epsilon\to0$. The conclusion is
that Eq.~\eqref{eq:(4.28)} is the unique contribution of the flow-line loop
diagrams for~$\epsilon\to0$.

By similar reasoning, it can be confirmed that Eq.~\eqref{eq:(4.28)} is the
unique place in which an apparent $O(\epsilon)$ term in the action contributes
in the $\epsilon\to0$ limit. The integration of the flow time of each vertex
eliminates one delta function and the singularity~$\delta(0)=1/\epsilon$ can
arise only from the flow-line loop diagrams, the case already considered above.
This observation justifies the Taylor expansion with respect to~$\epsilon$
and the neglect of the $O(\epsilon)$ terms besides that particular term
in~Eq.~\eqref{eq:(4.29)}.

Thus, we have observed that the perturbative expansions in the above two
systems are equivalent by a remarkable mechanism: A flow-line loop diagram in
the $(D+1)$-dimensional system, which is not generated in the perturbative
expansion of the original flow equation, reproduces the measure
term~\eqref{eq:(2.16)} which is absent in the original partition function of
the $(D+1)$-dimensional system, Eq.~\eqref{eq:(4.10)}. The mechanism is
remarkable, because an apparent $O(\epsilon)$ term in the action, i.e.,
$\mathcal{E}$~\eqref{eq:(4.13)}, plays the crucial role through the flow-line
loop.

Now, having established the equivalence between the $(D+1)$-dimensional field
theory~\eqref{eq:(4.10)}--\eqref{eq:(4.13)} and the flowed system
in~Sect.~\ref{sec:2}, we are ready to prove the UV finiteness of the gradient
flow in~Sect.~\ref{sec:2}.

\section{Proof of the renormalizability of the gradient flow}
\label{sec:5}
In this section, on the basis of the $(D+1)$-dimensional field theory in the
preceding section, we show that any correlation function of the flowed
$N$-vector field in terms of the renormalized coupling is UV finite, without
the wave function renormalization.

\subsection{Residual non-linear symmetry}
We first note that even with the gauge fixing~\eqref{eq:(4.7)}, there remains
a residual symmetry that is a particular combination of the global $O(N)$
symmetry~\eqref{eq:(4.4)} and the local symmetries~\eqref{eq:(4.5)}. It is
given by the requirement that it does not affect the gauge fixing conditions.
That is,
\begin{align}
   \delta\xi^N(x)&=\epsilon^{Nk}\xi^k(x)+g(x)\overline{n}^N(x)=0,
\label{eq:(5.1)}\\
   \delta\lambda^N(t,x)&=\epsilon^{Nk}\lambda^k(t,x)+h(t,x)n^N(t,x)=0.
\label{eq:(5.2)}
\end{align}
From these, we have
\begin{equation}
   g(x)=\frac{\epsilon^{kN}\xi^k(x)}{\overline{n}^N(x)},\qquad
   h(t,x)=\frac{\epsilon^{kN}\lambda^k(t,x)}{n^N(t,x)}.
\label{eq:(5.3)}
\end{equation}
Under this residual symmetry, other field components transform as
\begin{align}
   \delta n^i(x)&=\epsilon^{ij}n^j(x),&
   \delta\xi^k(x)
   &=\epsilon^{kl}\xi^l(x)
   +\epsilon^{lN}\xi^l(x)\frac{\overline{n}^k(x)}{\overline{n}^N(x)},
\notag\\
   \delta n^i(t,x)&=\epsilon^{ij}n^j(t,x),&
   \delta\lambda^k(t,x)
   &=\epsilon^{kl}\lambda^l(t,x)
   +\epsilon^{lN}\lambda^l(t,x)\frac{n^k(t,x)}{n^N(t,x)},
\label{eq:(5.4)}
\end{align}
where indices $k$ and~$l$ run over only from $1$ to~$N-1$.

The interesting part in the above residual symmetry is the $O(N)/O(N-1)$ part
corresponding to the choice of parameters~$\epsilon^{kl}=0$. Writing
$\epsilon^k\equiv\epsilon^{kN}$, it induces the following non-linear
transformations
\begin{align}
   \delta\pi^k(x)&=\epsilon^k\sqrt{1-\pi(x)^2},&
   \delta\xi^k(x)
   &=\epsilon^l\xi^l(x)\frac{\overline{\pi}^k(x)}{\overline{\sqrt{1-\pi(x)^2}}},
\notag\\
   \delta\pi^k(t,x)&=\epsilon^k\sqrt{1-\pi(t,x)^2},&
   \delta\lambda^k(t,x)
   &=\epsilon^l\lambda^l(t,x)\frac{\pi^k(t,x)}{\sqrt{1-\pi(t,x)^2}}.
\label{eq:(5.5)}
\end{align}
It can be directly confirmed that the integration measure and the action
in~Eqs.~\eqref{eq:(4.10)}--\eqref{eq:(4.13)} are invariant under this
non-linear transformation; this is expected, because the original partition
function with the discrete flow time, Eq.~\eqref{eq:(4.1)}
with~Eq.~\eqref{eq:(4.2)}, is invariant under Eqs.~\eqref{eq:(4.4)}
and~\eqref{eq:(4.5)}.

\subsection{Ward--Takahashi relation or the Zinn-Justin equation}
We can express the invariance of the system under the non-linear
transformation~\eqref{eq:(5.5)} as an identity for the generating functional of
1PI correlation functions. First, we introduce the source terms for elementary
fields,
\begin{equation}
   S_J\equiv-\int\mathrm{d}^Dx\,\left[J_\pi^k(x)\pi^k(x)\right]
   -\epsilon\sum_{t=0}^\infty\int\mathrm{d}^Dx\,
   \left[J_\pi^k(t,x)\pi^k(t,x)+J_\lambda^k(t,x)\lambda^k(t,x)\right],
\label{eq:(5.6)}
\end{equation}
\emph{except\/} for the Lagrange multiplier field~$\xi^k(x)$. It turns out that
this omission of the $\xi$-source greatly simplifies the discussion of the
renormalization. This implies that we omit correlation functions including
$\xi^k(x)$ from our consideration. However, since the $\xi$-field appears only
in the quadratic (i.e., free) part of the action~$S$ only linearly, if 1PI
correlation functions of other elementary fields turn out to be UV finite after
renormalization, any correlation functions including the elementary $\xi$-field
are also UV finite. Hence nothing is lost by the omission of the $\xi$-source
for our present purpose.

To write down the Ward--Takahashi relation associated with the
symmetry~\eqref{eq:(5.5)}, we also supplement additional terms to the action,
as
\begin{equation}
   S_{\text{tot}}=S+S_H+S_K,
\label{eq:(5.7)}
\end{equation}
where
\begin{align}
   S_H&\equiv-\int\mathrm{d}^Dx\,H(x)\sqrt{1-\pi(x)^2}
   -\epsilon\sum_{t=0}^\infty\int\mathrm{d}^Dx\,H(t,x)\sqrt{1-\pi(t,x)^2},
\label{eq:(5.8)}\\
   S_K&\equiv
   -\epsilon\sum_{t=0}^\infty\int\mathrm{d}^Dx\,\sum_{n=1}^\infty
   K^{k,l_1\dots l_n}(t,x)\mathcal{O}^{k,l_1\dots l_n}(t,x),
\label{eq:(5.9)}
\end{align}
and
\begin{equation}
   \mathcal{O}^{k,l_1\dots l_n}(t,x)
   \equiv
   \lambda^k(t,x)\frac{\pi^{l_1}(t,x)}{\sqrt{1-\pi(t,x)^2}}\dotsb
   \frac{\pi^{l_n}(t,x)}{\sqrt{1-\pi(t,x)^2}},
\label{eq:(5.10)}
\end{equation}
where the source $K^{k,l_1\dots l_n}(t,x)$ is symmetric in
indices~$(l_1,\dots,l_n)$ by definition.

We now consider the variation of integration variables of the form
of~Eq.~\eqref{eq:(5.5)} in the partition function:
\begin{align}
   \mathcal{Z}''&=
   \int
   \left[\prod_{k=1}^{N-1}\mathcal{D}\xi^k\right]
   \left[\prod_{k=1}^{N-1}\mathcal{D}\pi^k\right]
\notag\\
   &\qquad{}\times
   \left[\prod_{t=0}^\infty\prod_{k=1}^{N-1}\mathcal{D}\lambda^k(t)\right]
   \left[\prod_{t=0}^\infty\prod_{k=1}^{N-1}\mathcal{D}\pi^k(t)\right]\,
   \prod_x\frac{\overline{\sqrt{1-\pi(x)^2}}}{\sqrt{1-\pi(x)^2}}\,
   \mathrm{e}^{-S_{\text{tot}}-S_J}.
\label{eq:(5.11)}
\end{align}
We note
\begin{equation}
   \delta\sqrt{1-\pi(x)^2}=-\epsilon^m\pi^m(x),\qquad
   \delta\sqrt{1-\pi(t,x)^2}=-\epsilon^m\pi^m(t,x),
\label{eq:(5.12)}
\end{equation}
and
\begin{equation}
   \delta\mathcal{O}^{k,l_1\dots l_n}(t,x)
   =\epsilon^m\left[\mathcal{O}^{m,kl_1\dots l_n}(t,x)
   +n\mathcal{O}^{k,ml_1\dots l_n}(t,x)\right]
   +\sum_{i=1}^n\epsilon^{l_i}
   \mathcal{O}^{k,l_1\dotsc\not{l}_i\dots l_n}(t,x).
\label{eq:(5.13)}
\end{equation}
Then, by the standard argument, the invariance of the integration measure and
of~$S$ imply that the generating functional of 1PI functions, defined by the
Legendre transformation,
\begin{equation}
   {\mit\Gamma}\equiv-\ln\mathcal{Z}''
   +\int\mathrm{d}^Dx\,\left[J_\pi^k(x)\pi^k(x)
   \right]
   +\epsilon\sum_{t=0}^\infty\int\mathrm{d}^Dx\,
   \left[J_\pi^k(t,x)\pi^k(t,x)
   +J_\lambda^k(t,x)\lambda^k(t,x)\right],
\label{eq:(5.14)}
\end{equation}
where $\pi^k(x)$, $\pi^k(t,x)$, and~$\lambda^k(t,x)$ denote expectation values
of elementary fields, follows an identity
\begin{align}
   &\int\mathrm{d}^Dx\,
   \frac{\delta{\mit\Gamma}}{\delta\pi^m(x)}
   \frac{\delta{\mit\Gamma}}{\delta H(x)}
   +\int_0^\infty\mathrm{d}t\int\mathrm{d}^Dx\,
   \left[
   \frac{\delta{\mit\Gamma}}{\delta\pi^m(t,x)}
   \frac{\delta{\mit\Gamma}}{\delta H(t,x)}
   +\frac{\delta{\mit\Gamma}}{\delta\lambda^k(t,x)}
   \frac{\delta{\mit\Gamma}}{\delta K^{m,k}(t,x)}
   \right]
\notag\\
   &\qquad{}
   +\int\mathrm{d}^Dx\,H(x)\pi^m(x)
   +\int_0^\infty\mathrm{d}t\int\mathrm{d}^Dx\,H(t,x)\pi^m(t,x)
\notag\\
   &\qquad{}+\int_0^\infty\mathrm{d}t\int\mathrm{d}^Dx\,
   K^{k,l}(t,x)
   \left[
   \frac{\delta{\mit\Gamma}}{\delta K^{m,kl}(t,x)}
   +\frac{\delta{\mit\Gamma}}{\delta K^{k,ml}(t,x)}
   -\delta^{lm}\lambda^k(t,x)
   \right]
\notag\\
   &\qquad{}
   +\int_0^\infty\mathrm{d}t\int\mathrm{d}^Dx\,
   \sum_{n=2}^\infty K^{k,l_1\dots l_n}(t,x)
\notag\\
   &\qquad\qquad{}
   \times
   \left[
   \frac{\delta{\mit\Gamma}}{\delta K^{m,kl_1\dots l_n}(t,x)}
   +n
   \frac{\delta{\mit\Gamma}}{\delta K^{k,ml_1\dots l_n}(t,x)}
   +\sum_{i=1}^n\delta^{l_im}
   \frac{\delta{\mit\Gamma}}{\delta K^{k,l_1\dotsc\not{l}_i\dots l_n}(t,x)}
   \right]=0.
\label{eq:(5.15)}
\end{align}
In writing down this identity, we have taken the continuum flow-time
limit~$\epsilon\to0$. This is justified because we have observed that the
symmetry~\eqref{eq:(5.5)} is preserved by the flow-time discretization. Also,
we have observed that the $(D+1)$-dimensional
system~Eqs.~\eqref{eq:(4.10)}--\eqref{eq:(4.13)} with~$\epsilon\to0$ reproduces
the perturbative expansion of the flow equation. Thus we can study the
renormalizability of the flowed system in~Sect.~\ref{sec:2} by using the
identity~\eqref{eq:(5.15)}.

\subsection{Structure of the renormalization}
Our statement of the renormalizability is that the 1PI generating
functional~$\mit\Gamma$ can be made UV finite in terms of renormalized
quantities, by appropriately choosing the constants $Z$ and~$Z_3$ in
\begin{equation}
   g_0^2\equiv\mu^\epsilon g^2Z,\qquad\pi^k(x)\equiv Z_3^{1/2}\pi_R^k(x),
   \qquad H(x)\equiv Z_3^{-1/2}H_R(x)
\label{eq:(5.16)}
\end{equation}
order by order in perturbation theory. In particular, we claim that the flowed
or ``bulk'' fields, $\pi^k(t,x)$ and~$\lambda^k(t,x)$, do not require
multiplicative renormalization.

Our argument proceeds by mathematical induction based on the loop expansion. We
set
\begin{equation}
   {\mit\Gamma}=\sum_{\ell=0}^\infty{\mit\Gamma}^{(\ell)},
\label{eq:(5.17)}
\end{equation}
where ${\mit\Gamma}^{(\ell)}$ is the generating functional in the $\ell$~th loop
order. The above assertion is certainly true for~$\ell=0$ (tree-level
approximation) for which $Z=Z_3=1$ is sufficient. Then suppose that, in
perturbation theory with renormalized quantities fixed, the constants $Z$
and~$Z_3$ in~Eq.~\eqref{eq:(5.16)} can be chosen so that ${\mit\Gamma}^{(0)}$,
\dots, ${\mit\Gamma}^{(\ell)}$, are UV finite in terms of renormalized
quantities. Then consider the $(\ell+1)$~th loop order calculation on the basis
of the above chosen $Z$ and~$Z_3$. Since $Z$ and~$Z_3$ have already been chosen
so that ${\mit\Gamma}^{(0)}$, \dots, ${\mit\Gamma}^{(\ell)}$ are finite, by
considering UV-divergent part of the identity~\eqref{eq:(5.15)} in the
$(\ell+1)$~th loop order, we have
\begin{equation}
   {\mit\Gamma}^{(0)}\ast{\mit\Gamma}^{(\ell+1)\text{div}}=0,
\label{eq:(5.18)}
\end{equation}
where ${\mit\Gamma}^{(\ell+1)\text{div}}$ denotes UV-divergent part
of~${\mit\Gamma}^{(\ell+1)}$ and
\begin{align}
   {\mit\Gamma}^{(0)}\ast
   &\equiv
   \int\mathrm{d}^Dx\,
   \left[
   \frac{\delta{\mit\Gamma}^{(0)}}{\delta\pi_R^m(x)}
   \frac{\delta}{\delta H_R(x)}
   +\frac{\delta{\mit\Gamma}^{(0)}}{\delta H_R(x)}
   \frac{\delta}{\delta\pi_R^m(x)}
   \right]
\notag\\
   &\qquad{}
   +\int_0^\infty\mathrm{d}t\int\mathrm{d}^Dx\,
   \left[
   \frac{\delta{\mit\Gamma}^{(0)}}{\delta\pi^m(t,x)}
   \frac{\delta}{\delta H(t,x)}
   +\frac{\delta{\mit\Gamma}^{(0)}}{\delta H(t,x)}
   \frac{\delta}{\delta\pi^m(t,x)}
   \right]
\notag\\
   &\qquad{}
   +\int_0^\infty\mathrm{d}t\int\mathrm{d}^Dx\,
   \left[
   \frac{\delta{\mit\Gamma}^{(0)}}{\delta\lambda^k(t,x)}
   \frac{\delta}{\delta K^{m,k}(t,x)}
   +\frac{\delta{\mit\Gamma}^{(0)}}{\delta K^{m,k}(t,x)}
   \frac{\delta}{\delta\lambda^k(t,x)}
   \right]
\notag\\
   &\qquad{}+\int_0^\infty\mathrm{d}t\int\mathrm{d}^Dx\,
   K^{k,l}(t,x)
   \left[
   \frac{\delta}{\delta K^{m,kl}(t,x)}
   +\frac{\delta}{\delta K^{k,ml}(t,x)}
   \right]
\notag\\
   &\qquad{}
   +\int_0^\infty\mathrm{d}t\int\mathrm{d}^Dx\,
   \sum_{n=2}^\infty K^{k,l_1\dots l_n}(t,x)
\notag\\
   &\qquad\qquad{}
   \times
   \left[
   \frac{\delta}{\delta K^{m,kl_1\dots l_n}(t,x)}
   +n
   \frac{\delta}{\delta K^{k,ml_1\dots l_n}(t,x)}
   +\sum_{i=1}^n\delta^{l_im}
   \frac{\delta}{\delta K^{k,l_1\dotsc\not{l}_i\dots l_n}(t,x)}
   \right],
\label{eq:(5.19)}
\end{align}
and
\begin{equation}
   {\mit\Gamma}^{(0)}\equiv\left.S_{\text{tot}}\right|_{Z=Z_3=1}.
\label{eq:(5.20)}
\end{equation}

We next study the most general form of the divergent
part~${\mit\Gamma}^{(\ell+1)\text{div}}$. First of all, by a general theorem, the
divergent part must be an integral of a local polynomial of fields and their
derivatives. We then note that there is no divergence corresponding to a local
term in the ``bulk'' $t>0$, a term that is written as an
$\int_0^\infty\mathrm{d}t\,\int\mathrm{d}^Dx$ integral of a local polynomial of
fields and their derivatives: As we explained in detail in~Sect~\ref{sec:4.4},
there is no loop diagram consisting solely of the ``flow-line''
$\pi\lambda$-propagator~\eqref{eq:(4.23)}, other than the diagram
in~Fig.~\ref{fig:10}, which reduces to the measure term at the boundary~$t=0$,
Eq.~\eqref{eq:(2.16)}. Then, since the $\pi\pi$-propagator~\eqref{eq:(4.22)}
possesses the Gaussian damping factor~$\mathrm{e}^{-(t+s)p^2}$
(for~$\epsilon\to0$), any loop diagram in which the flow times of the vertices
(they must be the same for the divergent part) are positive is UV finite.
Therefore, there is no divergence that is written as the bulk integral.

Any divergent part is thus written as the integral on the boundary~$t=0$.
Noting that for $D=2$ the fields $\pi_R^k(x)$ and~$\pi^k(t,x)$ possess the the
mass dimension~$0$, $H_R(x)$, $\lambda^k(t,x)$, and~$K^{k,l_1\dots l_n}(t,x)$
possess~$2$, and $H(t,x)$ possesses~$4$, the most general possible form of the
divergent part is
\begin{align}
   {\mit\Gamma}^{(\ell+1)\text{div}}
   &=\int\mathrm{d}^Dx\,\Bigl[
   B(\pi_R(x),\partial_\mu\pi_R(x))
   +H_R(x)C(\pi_R(x))
\notag\\
   &\qquad\qquad\qquad{}
   +\lambda^k(0,x)D^k(\pi_R(x))
   +\sum_{n=1}^\infty K^{k,l_1\dots l_n}(0,x)E^{k,l_1\dots l_n}(\pi_R(x))
   \Bigr],
\label{eq:(5.21)}
\end{align}
where $B$ contains at most two derivatives and $E^{k,l_1\dots l_n}$ is symmetric
in indices~$(l_1,\dots,l_n)$. Note that we have not included the flow field at
zero flow time, $\pi^k(0,x)$, in the possible form of the divergent
part~\eqref{eq:(5.21)}. The redundancy to use this field variable in addition
to~$\pi_R^k(x)$ follows from the relation
\begin{equation}
   \pi^k(0,x)=\pi^k(x),
\label{eq:(5.22)}
\end{equation}
i.e., the expectation value of the variation of the action with respect to
the $\xi$-field. Note that here the field variables denote the expectation
values in the presence of source fields and \emph{not\/} the integration
variables in the functional integral. This identity shows that as the arguments
of the 1PI generating functional, the variables $\pi^k(0,x)$ and~$\pi^k(x)$
cannot be independent, because they cannot take different values for any
configuration of the source fields.

We note also that the combination
\begin{equation}
   \int\mathrm{d}^Dx\,\partial_t\pi^k(0,x)F^k(\pi_R(x))
\label{eq:(5.23)}
\end{equation}
does not appear in~Eq.~\eqref{eq:(5.21)}: An external $\pi^k(t,x)$ line in a
1PI diagram can arise only from a flow vertex that inevitably contains the
Lagrange multiplier field~$\lambda^k(t,x)$. Since there is no flow-line loop
(other than the diagram in~Fig.~\ref{fig:10} which reduces to a boundary term),
the flow-line propagator starting from~$\lambda^k(t,x)$ can end only at
\emph{another\/} flow vertex that contains another $\lambda^k(s,x)$. This shows
that any 1PI diagram containing $\pi^k(t,x)$ must accomplish at least
one~$\lambda$. The combination~\eqref{eq:(5.23)} does not match this rule.

Now, having obtained the general form of the divergent part,
Eq.~\eqref{eq:(5.21)}, we examine the implication of the
identity~\eqref{eq:(5.18)} with~Eq.~\eqref{eq:(5.19)}.

First of all, examining the coefficient of~$\partial_t\pi^k(0,x)$
in~Eq.~\eqref{eq:(5.18)} that arises
from~$\delta{\mit\Gamma}^{(0)}/\delta\lambda^k(0,x)$ in~Eq.~\eqref{eq:(5.19)},
we have
\begin{equation}
   E^{m,k}=0.
\label{eq:(5.24)}
\end{equation}
Then, from various terms in~Eq.~\eqref{eq:(5.18)}, we have
\begin{align}
   &\frac{\partial C}{\partial\pi_R^m(x)}=\frac{\pi_R^m(x)}{1-\pi_R(x)^2}C,
\label{eq:(5.25)}\\
   &\int\mathrm{d}^Dx\,
   \sqrt{1-\pi_R(x)^2}\frac{\delta}{\delta\pi_R^m(x)}\int\mathrm{d}^Dx\,B
\notag\\
   &\qquad{}=\int\mathrm{d}^Dx\,
   \frac{1}{\mu^\epsilon g^2}\left[-\partial_\mu\partial_\mu\pi_R^m(x)
   +\frac{\pi_R^m(x)}{\sqrt{1-\pi_R(x)^2}}
   \partial_\mu\partial_\mu\sqrt{1-\pi_R(x)^2}\right]C,
\label{eq:(5.26)}\\
   &\frac{\partial D^k}{\partial\pi_R^m(x)}
   +\delta^{mk}\frac{\pi_R^l(x)}{1-\pi_R(x)^2}D^l
   =0,
\label{eq:(5.27)}
\end{align}
and
\begin{align}
   &E^{m,kl}+E^{k,ml}=
   \sqrt{1-\pi_R(x)^2}\,\frac{\partial E^{k,l}}{\partial\pi_R^m(x)},
\label{eq:(5.28)}\\
   &E^{m,kl_1\dots l_n}+nE^{k,ml_1\dots l_n}
   =-\sum_{i=1}^n\delta^{l_im}E^{k,l_1\dotsc\not{l}_i\dots l_n}
   +\sqrt{1-\pi_R(x)^2}\,
   \frac{\partial E^{k,l_1\dots l_n}}{\partial\pi_R^m(x)},
   \qquad n\geq2.
\label{eq:(5.29)}
\end{align}

The above conditions for $B$ and~$C$, Eqs.~\eqref{eq:(5.25)}
and~\eqref{eq:(5.26)}, are completely identical to the conditions on the
divergent part in the original 2D $O(N)$ non-linear sigma
model~\cite{Brezin:1976ap}. The general solution to these is given
by~\cite{Brezin:1976ap}
\begin{align}
   C&=-\frac{1}{2}\delta Z_3\frac{1}{\sqrt{1-\pi_R(x)^2}},
\label{eq:(5.30)}\\
   B&=\delta Z\frac{1}{2\mu^\epsilon g^2}
   \left\{
   \left[\partial_\mu\pi_R(x)\right]^2
   +\left[\partial_\mu\sqrt{1-\pi_R(x)^2}\right]^2\right\}
\notag\\
   &\qquad{}
   -\delta Z_3\frac{1}{2\mu^\epsilon g^2}
   \left\{
   \left[\partial_\mu\pi_R(x)\right]^2
   -\partial_\mu\sqrt{1-\pi_R(x)^2}
   \partial_\mu\frac{\pi_R(x)^2}{\sqrt{1-\pi_R(x)^2}}\right\},
\label{eq:(5.31)}
\end{align}
where $\delta Z$ and~$\delta Z_3$ are constants.

Next, from the linearly realized $O(N-1)$ symmetry (that is preserved in our
all steps), one can set $D^k=\pi_R^k(x)d(\pi_R(x)^2)$. Then
Eq.~\eqref{eq:(5.27)} immediately shows that~$d=0$ and
\begin{equation}
   D^k=0.
\label{eq:(5.32)}
\end{equation}

Next, from Eqs.~\eqref{eq:(5.24)} and~\eqref{eq:(5.28)}, and the fact that
$E^{m,kl}$ is symmetric under the exchange~$k\leftrightarrow l$, we have
\begin{equation}
   E^{m,kl}=-E^{k,ml}=-E^{k,lm}=+E^{l,km}=+E^{l,mk}=-E^{m,lk}=-E^{m,kl}=0.
\label{eq:(5.33)}
\end{equation}

Finally, we note that, if the right-hand side of~Eq.~\eqref{eq:(5.29)}
vanishes, then
\begin{equation}
   E^{m,kl_1\dots l_n}=-nE^{k,ml_1\dots l_n}=+n^2E^{m,kl_1\dots l_n},
\label{eq:(5.34)}
\end{equation}
and thus
\begin{equation}
   E^{m,kl_1\dots l_n}=0,\qquad n\geq2.
\label{eq:(5.35)}
\end{equation}
This is actually the case by mathematical induction because the right-hand side
of~Eq.~\eqref{eq:(5.29)} vanishes for~$n=2$ from~Eqs.~\eqref{eq:(5.24)}
and~\eqref{eq:(5.33)} and then for~$n=3$ again from~Eq.~\eqref{eq:(5.33)}; we
see that $E^{k,l_1\dots l_n}=0$ for all~$n\geq1$.

In summary, we observed that possible divergent part in the present system is
given by~Eq.~\eqref{eq:(5.21)} with~Eqs.~\eqref{eq:(5.30)}
and~\eqref{eq:(5.31)} and $D^k=E^{k,i_1\dots l_n}=0$. This is identical to the
divergent part in the 2D $O(N)$ non-linear sigma model. One can see that the
divergent part~\eqref{eq:(5.21)} with~Eqs.~\eqref{eq:(5.30)}
and~\eqref{eq:(5.31)} is canceled by the variation of the total action
$S_{\text{tot}}$~\eqref{eq:(5.7)} under the change of the renormalization
constants in~Eq.~\eqref{eq:(5.16)} by $(\ell+1)$~th loop order quantities:
\begin{equation}
   Z\to Z+\delta Z,\qquad Z_3\to Z_3+\delta Z_3.
\label{eq:(5.36)}
\end{equation}
The 1PI generating functional in the $(\ell+1)$~th loop order,
${\mit\Gamma}^{(\ell+1)}$, is thus made UV finite. This completes the
mathematical induction for the renormalizability. In particular, we showed that
\emph{there is no need of the wave function renormalization for the flowed
fields}, $\pi^k(t,x)$ and~$\lambda^k(t,x)$.

We have shown that any correlation function of the flowed fields is UV finite
under the conventional parameter renormalization, without multiplicative wave
function renormalization. Then, it is easy to see that, because of Gaussian
damping factors in propagators, this UV finiteness holds even when some
spacetime coordinates of the correlation function coincide, i.e., even in the
equal-point limit, as long as all flow-time coordinates of flowed fields are
strictly positive.\footnote{Again, the absence of the flow-line loop diagram is
crucial for this finiteness.} The local product of any number of flowed fields
does not contain UV divergences. This robust UV finiteness, that makes the
construction of renormalized composite operators straightforward, is the key
property in application of the gradient flow in lattice field theory.

In renormalized perturbation theory, one uses the propagators and the vertices
in terms of renormalized parameters and renormalized fields. This renormalized
Feynman rule is obtained by making the substitution~\eqref{eq:(5.16)} in the
action~\eqref{eq:(4.12)}. The part including the $\xi$-field becomes
\begin{equation}
   -i\int\mathrm{d}^Dx\,\xi^k(x)
   \left[\pi^k(0,x)-\pi_R^k(x)\right]
   +i\int\mathrm{d}^Dx\,(Z_3^{1/2}-1)
   \xi^k(x)\pi_R^k(x),
\label{eq:(5.37)}
\end{equation}
and the second term is regarded as the perturbation. In this renormalized
perturbation theory, from the first term, the free propagator is given by
\begin{equation}
   \left\langle\pi^k(t,x)\pi^l(s,y)\right\rangle_0
   =\mu^\epsilon g^2\delta^{kl}\int_p\mathrm{e}^{ip(x-y)}
   \frac{\mathrm{e}^{-(t+s)p^2}}{p^2},
\label{eq:(5.38)}
\end{equation}
while the second term is regarded as a counterterm. In this way, we can also
use Eq.~\eqref{eq:(5.38)} for~$\pi_R^k(x)$ by identifying
$\pi_R^k(x)=\pi^k(0,x)$. As the $\pi\xi$-propagator~\eqref{eq:(4.24)} shows,
the second term in~Eq.~\eqref{eq:(5.37)} acts as a two-point vertex at the
boundary~$t=0$ that connects between $\pi^k(t,x)$ and~$\pi_R^l(y)$. This
counterterm thus plays the same role as the boundary
counterterm~$\Delta S_{\text{bc}}$ in the gauge theory (Sect.~7.1
of~Ref.~\cite{Luscher:2011bx}).

Finally, the IR-regulating mass term~\eqref{eq:(2.8)} can readily be
incorporated in the above argument by the substitution
\begin{equation}
   H(x)\to H(x)+\frac{m_0^2}{g_0^2}.
\label{eq:(5.39)}
\end{equation}
In particular, from~Eq.~\eqref{eq:(5.16)}, we see that the generating
functional becomes UV finite in terms of
\begin{equation}
   Z_3^{1/2}\left[H(x)+\frac{m_0^2}{g_0^2}\right]
   =H_R(x)+\frac{1}{\mu^\epsilon g^2}\frac{Z_3^{1/2}m_0^2}{Z}.
\label{eq:(5.40)}
\end{equation}
This shows that the mass parameter is renormalized as~$m_0^2=(Z/Z_3^{1/2})m^2$,
as we already noted in~Eq.~\eqref{eq:(3.6)}.

\section{Lattice energy--momentum tensor}
\label{sec:6}

In the preceding section, we have shown that any local product (the composite
operator) of the bare flowed $N$-vector field becomes UV finite under the
conventional parameter renormalization in the 2D $O(N)$ non-linear sigma model.
As application of this fact, in the present section, we consider the
construction of the energy--momentum tensor, the Noether current associated
with the translational invariance, in a lattice formulation of the non-linear
sigma model. The idea is the same as that in~Refs.~\cite{Suzuki:2013gza}
and~\cite{Makino:2014taa}: Since lattice regularization explicitly breaks the
translational invariance, the construction of the energy--momentum tensor is
awkward. Instead of considering this construction directly, we construct a
composite operator of the flowed field which, under dimensional regularization,
becomes the energy--momentum tensor. Since dimensional regularization preserves
the translational invariance, the description of the energy--momentum tensor
that fulfills the correct Ward--Takahashi relation is straightforward. On the
other hand, since the composite operator of the flowed field is UV finite under
the parameter renormalization, it must become independent of the regularization
in the limit that the regulator is removed (after the renormalization, as long
as the same renormalization conditions are adopted). In this way, low-energy
correlation functions of the energy--momentum tensor may be computed by using
lattice regularization. This construction in~Ref.~\cite{Suzuki:2013gza} has
been applied to the thermodynamics of quenched QCD
in~Ref.~\cite{Asakawa:2013laa} and promising results have been obtained.

\subsection{Energy--momentum tensor with dimensional regularization}
The energy--momentum tensor~$T_{\mu\nu}(x)$ can be obtained from the variation
of the action~\eqref{eq:(2.1)} under the infinitesimal translation with a
localized parameter,
\begin{equation}
   \delta n^i(x)=\xi_\mu(x)\partial_\mu n^i(x),
\label{eq:(6.1)}
\end{equation}
as
\begin{equation}
   \delta S=-\int\mathrm{d}^Dx\,\xi_\nu(x)\partial_\mu T_{\mu\nu}(x).
\label{eq:(6.2)}
\end{equation}
The explicit form is given by
\begin{equation}
   T_{\mu\nu}(x)=\frac{1}{g_0^2}
   \left[
   \partial_\mu n^i(x)\partial_\nu n^i(x)
   -\frac{1}{2}\delta_{\mu\nu}\partial_\rho n^i(x)\partial_\rho n^i(x)
   \right].
\label{eq:(6.3)}
\end{equation}
Assuming that we are using dimensional regularization, which preserves the
translational invariance, the above classical expression as it stands fulfills
the correct Ward--Takahashi relation associated with the translational
invariance:
\begin{equation}
   \left\langle\mathcal{O}_{\text{ext}}
   \int_{\mathcal{D}}
   \mathrm{d}^Dx\,\partial_\mu
   \left\{T_{\mu\nu}\right\}_R(x)\,\mathcal{O}_{\text{int}}\right\rangle
   =-\left\langle\mathcal{O}_{\text{ext}}\,\partial_\nu\mathcal{O}_{\text{int}}
   \right\rangle.
\label{eq:(6.4)}
\end{equation}
In this expression, $\mathcal{D}$ is a bounded integration region,
$\mathcal{O}_{\text{ext}}$ is an operator outside the region~$\mathcal{D}$, and
$\mathcal{O}_{\text{int}}$ is an operator inside the region. We defined the
renormalized energy--momentum tensor by subtracting the vacuum expectation
value, $\{T_{\mu\nu}\}_R(x)\equiv T_{\mu\nu}(x)-\langle T_{\mu\nu}(x)\rangle$. The
Ward--Takahashi relation ensures that the bare quantity $T_{\mu\nu}(x)$ is not
multiplicatively renormalized.

Although naively the energy--momentum tensor~\eqref{eq:(6.3)} is traceless
for~$D=2$, UV divergences in the composite
operator~$(1/g_0^2)\partial_\rho n^i(x)\partial_\rho n^i(x)$ being proportional
to~$1/\epsilon$ makes this expectation invalid even for~$\epsilon\to0$.
Instead, we have the the trace anomaly,
\begin{equation}
   \delta_{\mu\nu}\left\{T_{\mu\nu}\right\}_R(x)
   =-\frac{\beta}{g^3}
   \left\{\partial_\rho n^i\partial_\rho n^i\right\}_R(x),
\label{eq:(6.5)}
\end{equation}
where the $\text{MS}$ scheme is assumed in the renormalized composite operator
in the right-hand side and the coefficient is given by the $\beta$ function,
\begin{equation}
   \beta\equiv\left(\mu\frac{\partial}{\partial\mu}\right)_0g
   =-\frac{\epsilon}{2}g-g^3\sum_{k=0}^\infty b_kg^{2k}
\label{eq:(6.6)}
\end{equation}
(here, the derivative with respect to the renormalization scale~$\mu$ is taken
while bare quantities are kept fixed),
and~\cite{Hikami:1977vr,Hikami:1982ak,Bernreuther:1986js}
\begin{equation}
   b_0=\frac{1}{4\pi}(N-2),\qquad b_1=\frac{1}{(4\pi)^2}2(N-2),\qquad
   b_2=\frac{1}{(4\pi)^3}(N-2)(N+2),
\label{eq:(6.7)}
\end{equation}
and
\begin{equation}
   b_3=\frac{1}{(4\pi)^4}(N-2)
   \left[-\frac{2}{3}(N^2-22N+34)+12(N-3)\zeta(3)\right].
\label{eq:(6.8)}
\end{equation}

\subsection{Small flow-time expansion and the energy--momentum tensor}
We construct a composite operator of the flowed field which reduces to the 2D
composite operator~\eqref{eq:(6.3)} by using the small flow-time expansion
introduced in~Ref.~\cite{Luscher:2011bx}. For this, we take an $O(N)$-invariant
dimension-$2$ second-rank composite operator of the flowed field:
\begin{equation}
   \partial_\mu n^i(t,x)\partial_\nu n^i(t,x)
   =\partial_\mu\pi^k(t,x)\partial_\nu\pi^k(t,x)
   +\partial_\mu\sqrt{1-\pi(t,x)^2}\partial_\nu\sqrt{1-\pi(t,x)^2}.
\label{eq:(6.9)}
\end{equation}
According to the argument in~Ref.~\cite{Luscher:2011bx}, for~$t\to0$, this
composite operator of the flowed field can be expressed as a series of 2D local
operators with increasing mass dimensions, as
\begin{align}
   &\partial_\mu n^i(t,x)\partial_\nu n^i(t,x)
\notag\\
   &=\left\langle\partial_\mu n^i(t,x)\partial_\nu n^i(t,x)\right\rangle
\notag\\
   &\qquad{}
   +\zeta_{11}(t)\left[
   \partial_\mu n^i(x)\partial_\nu n^i(x)
   -\left\langle\partial_\mu n^i(x)\partial_\nu n^i(x)\right\rangle
   \right]
\notag\\
   &\qquad\qquad{}
   +\zeta_{12}(t)\left[
   \delta_{\mu\nu}\partial_\rho n^i(x)\partial_\rho n^i(x)
   -\left\langle\delta_{\mu\nu}\partial_\rho n^i(x)\partial_\rho n^i(x)
   \right\rangle
   \right]
   +O(t).
\label{eq:(6.10)}
\end{align}
Similarly, we have
\begin{align}
   &\partial_\rho n^i(t,x)\partial_\rho n^i(t,x)
\notag\\
   &=\left\langle\partial_\rho n^i(t,x)\partial_\rho n^i(t,x)\right\rangle
   +\zeta_{22}(t)
   \left[\partial_\rho n^i(x)\partial_\rho n^i(x)
   -\left\langle\partial_\rho n^i(x)\partial_\rho n^i(x)\right\rangle
   \right]+O(t).
\label{eq:(6.11)}
\end{align}

Inverting these relations with respect to the 2D operators and substituting
them into~Eq.~\eqref{eq:(6.3)}, we have
\begin{align}
   \left\{T_{\mu\nu}\right\}_R(x)
   &\equiv T_{\mu\nu}(x)-\left\langle T_{\mu\nu}(x)\right\rangle
\notag\\
   &=c_1(t)\left[
   \partial_\mu n^i(t,x)\partial_\nu n^i(t,x)
   -\frac{1}{2}\delta_{\mu\nu}\partial_\rho n^i(t,x)\partial_\rho n^i(t,x)
   \right]
\notag\\
   &\qquad{}+c_2(t)\left[
   \frac{1}{2}
   \delta_{\mu\nu}\partial_\rho n^i(t,x)\partial_\rho n^i(t,x)
   -\left\langle
   \frac{1}{2}
   \delta_{\mu\nu}\partial_\rho n^i(t,x)\partial_\rho n^i(t,x)
   \right\rangle
   \right]+O(t),
\label{eq:(6.12)}
\end{align}
where
\begin{align}
   c_1(t)&=\frac{1}{g_0^2}\zeta_{11}(t)^{-1},
\label{eq:(6.13)}
\\
   c_2(t)&=\frac{1}{g_0^2}
   \left\{\left[-2\zeta_{11}(t)^{-1}\zeta_{12}(t)-1\right]
   \zeta_{22}(t)^{-1}+\zeta_{11}(t)^{-1}\right\}.
\label{eq:(6.14)}
\end{align}
Hence, if we know the $t\to0$ behavior of the coefficients~$\zeta_{IJ}(t)$
in~Eqs.~\eqref{eq:(6.10)} and~\eqref{eq:(6.11)}, then the energy--momentum
tensor~\eqref{eq:(6.3)} can be obtained by the $t\to0$ limit of the right-hand
side of~Eq.~\eqref{eq:(6.12)}.

Thus, we are interested in the $t\to0$ behavior of the
coefficients~$\zeta_{IJ}(t)$. Since all the composite operators in the above
expansions are bare ones, by the standard renormalization group argument, the
expansion coefficients are independent of the renormalization scale~$q$, if
they are expressed in terms of the running parameter~$\Bar{g}(q)$. In
particular, we may take $q=1/\sqrt{8t}$. Then because of asymptotic freedom,
the running coupling behaves as~$\Bar{g}(1/\sqrt{8t})\to0$ for~$t\to0$
and~$\zeta_{IJ}(t)$ for~$t\to0$ can be evaluated by perturbation theory.

To find the coefficients~$\zeta_{IJ}(t)$ in~Eq.~\eqref{eq:(6.10)}, we consider
correlation functions,
\begin{equation}
   \left\langle\partial_\mu n^i(t,x)\partial_\nu n^i(t,x)
   \pi^k(y)\pi^l(z)\right\rangle
   \equiv g_0^4\delta^{kl}\int_{p,q}\,
   \frac{\mathrm{e}^{ip(x-y)}}{p^2+m_0^2}\frac{\mathrm{e}^{iq(x-z)}}{q^2+m_0^2}\,
   \mathcal{M}_{\mu\nu}(p,q;t)
\label{eq:(6.15)}
\end{equation}
and
\begin{equation}
   \left\langle\partial_\mu n^i(x)\partial_\nu n^i(x)
   \pi^k(y)\pi^l(z)\right\rangle
   \equiv g_0^4\delta^{kl}\int_{p,q}\,
   \frac{\mathrm{e}^{ip(x-y)}}{p^2+m_0^2}\frac{\mathrm{e}^{iq(x-z)}}{q^2+m_0^2}\,
   \mathcal{M}_{\mu\nu}(p,q),
\label{eq:(6.16)}
\end{equation}
and compute the coefficients of combinations,
\begin{equation}
   ip_\mu iq_\nu+iq_\mu ip_\nu,\qquad 2\delta_{\mu\nu}ip_\rho iq_\rho,
\label{eq:(6.17)}
\end{equation}
in tensors~$\mathcal{M}_{\mu\nu}(p,q;t)$ and~$\mathcal{M}_{\mu\nu}(p,q)$. Then,
we determine $\zeta_{IJ}(t)$ so that the relation~\eqref{eq:(6.10)} holds in
correlation functions in view of the combinations~\eqref{eq:(6.17)}.

\begin{figure}
\begin{center}
\includegraphics[width=3cm,clip]{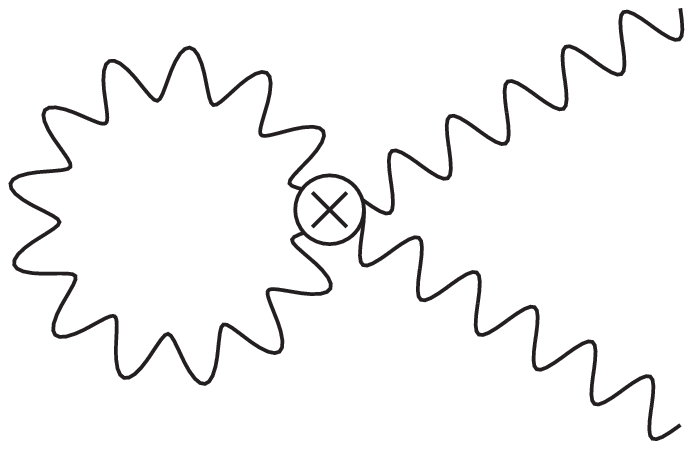}
\caption{Diagram 07: A one-loop diagram that contributes to Eq.~\eqref{eq:(6.15)}.}
\label{fig:12}
\end{center}
\end{figure}

\begin{figure}
\begin{center}
\includegraphics[width=3cm,clip]{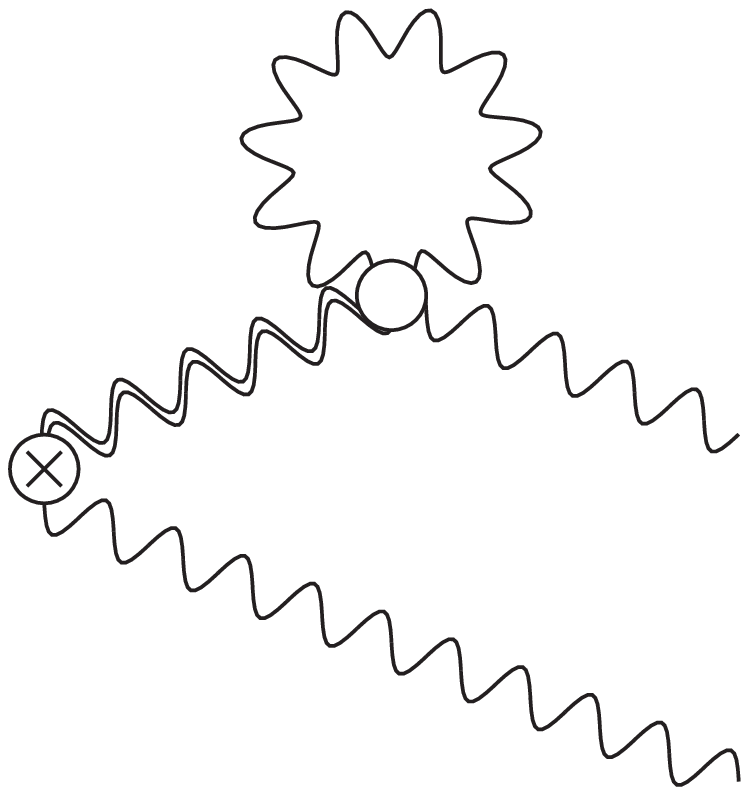}
\caption{Diagram 08: A one-loop diagram that contributes to Eq.~\eqref{eq:(6.15)}.}
\label{fig:13}
\end{center}
\end{figure}

\begin{figure}
\begin{center}
\includegraphics[width=3cm,clip]{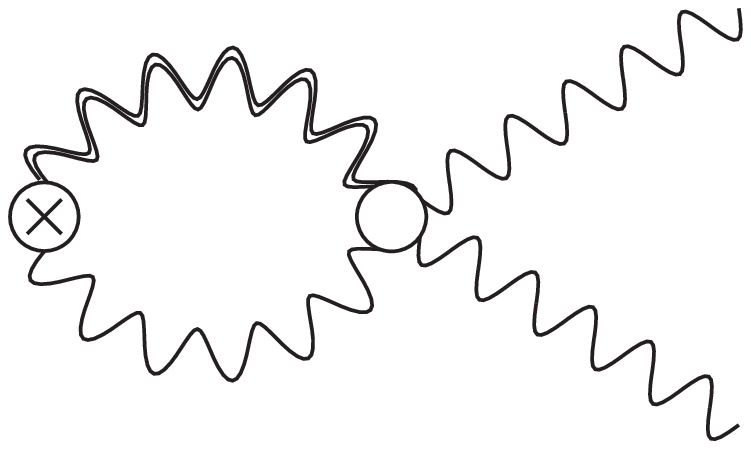}
\caption{Diagram 09: A one-loop diagram that contributes to Eq.~\eqref{eq:(6.15)}.}
\label{fig:14}
\end{center}
\end{figure}

\begin{figure}
\begin{center}
\includegraphics[width=3cm,clip]{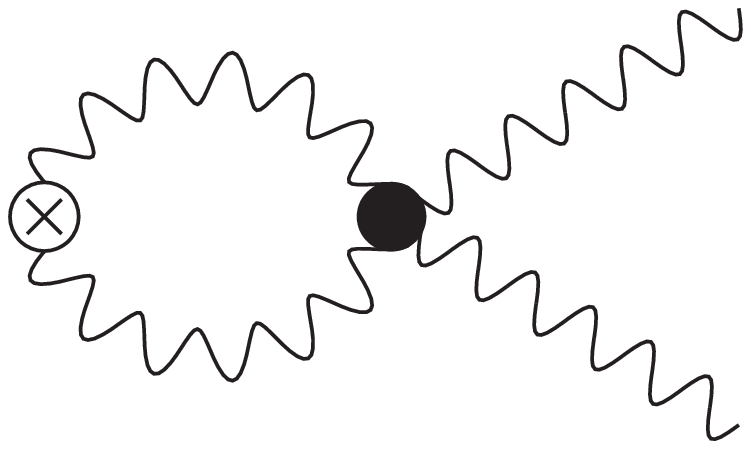}
\caption{Diagram 10: A one-loop diagram that contributes to Eq.~\eqref{eq:(6.15)}.}
\label{fig:15}
\end{center}
\end{figure}

Set $\zeta_{IJ}(t)=\zeta_{IJ}^{(0)}(t)+\zeta_{IJ}^{(1)}(t)+\dotsb$, where
superscripts denote the number of loops. In the tree level,
$\zeta_{IJ}^{(0)}(t)=\delta_{IJ}$. From this, it follows that the one-loop
correction $\zeta_{11}^{(1)}(t)$ ($\zeta_{12}^{(1)}(t)$) is given by the
difference of one-loop coefficients of the former (the latter) combination
in~Eq.~\eqref{eq:(6.17)} between Eqs.~\eqref{eq:(6.15)} and~\eqref{eq:(6.16)}.
In the one-loop level, for~Eq.~\eqref{eq:(6.15)}, we have four diagrams
in~Figs.~\ref{fig:12}--\ref{fig:15}.\footnote{We neglect standard one-particle
irreducible diagrams because they give rise to the same contributions
to~Eq.~\eqref{eq:(6.15)} and to~Eq.~\eqref{eq:(6.16)}.}
For~Eq.~\eqref{eq:(6.16)}, we have only two diagrams in~Figs.~\ref{fig:12}
and~\ref{fig:15}. We use dimensional regularization to regularize the UV
divergences and the mass term~\eqref{eq:(2.8)} to regularize the IR divergences.

%
\begin{table}
\caption{One-loop coefficients of the combinations~\eqref{eq:(6.17)}
in~$\mathcal{M}_{\mu\nu}(p,q;t)$ in~Eq.~\eqref{eq:(6.15)} in units
of~$g_0^2/(4\pi)$.}
\label{table:1}
\begin{center}
\renewcommand{\arraystretch}{2.2}
\setlength{\tabcolsep}{20pt}
\begin{tabular}{crr}
\toprule
diagram & \multicolumn{1}{c}{$ip_\mu iq_\nu+iq_\mu ip_\nu$} & \multicolumn{1}{c}{$2\delta_{\mu\nu}ip_\rho iq_\rho$} \\
\midrule
07  & $-\ln(2\mathrm{e}^{\gamma_\mathrm{E}}m_0^2t)$ & $0$ \\
08  & $(2N-2)\left[\dfrac{1}{\epsilon}+\dfrac{1}{2}\ln(8\pi t)\right]$ & $0$ \\
09  & $\dfrac{3}{4}$ & $\dfrac{1}{2}N-\dfrac{5}{8}$ \\
10  & $-\dfrac{5}{12}$ & $\dfrac{1}{2}(-N+2)\ln(2\mathrm{e}^{\gamma_\mathrm{E}}m_0^2t)-\dfrac{N}{2}+\dfrac{19}{24}$ \\
\bottomrule
\end{tabular}
\end{center}
\end{table}
\begin{table}
\caption{One-loop coefficients of the combinations~\eqref{eq:(6.17)}
in~$\mathcal{M}_{\mu\nu}(p,q)$ in~Eq.~\eqref{eq:(6.16)} in units
of~$g_0^2/(4\pi)$.}
\label{table:2}
\begin{center}
\renewcommand{\arraystretch}{2.2}
\setlength{\tabcolsep}{20pt}
\begin{tabular}{crr}
\toprule
diagram & \multicolumn{1}{c}{$ip_\mu iq_\nu+iq_\mu ip_\nu$} & \multicolumn{1}{c}{$2\delta_{\mu\nu}ip_\rho iq_\rho$} \\
\midrule
07 & $2\left[\dfrac{1}{\epsilon}-\dfrac{1}{2}\ln\left(\dfrac{\mathrm{e}^{\gamma_\mathrm{E}}m_0^2}{4\pi}\right)\right]$ & $0$ \\
10 & $\dfrac{1}{3}$
& $(N-2)\left[\dfrac{1}{\epsilon}-\dfrac{1}{2}\ln\left(\dfrac{\mathrm{e}^{\gamma_\mathrm{E}}m_0^2}{4\pi}\right)\right]+\dfrac{1}{6}$ \\
\bottomrule
\end{tabular}
\end{center}
\end{table}

The contribution of each diagram to~$\mathcal{M}_{\mu\nu}(p,q;t)$
in~Eq.~\eqref{eq:(6.15)} and to~$\mathcal{M}_{\mu\nu}(p,q)$
in~Eq.~\eqref{eq:(6.16)} is tabulated in~Tables~\ref{table:1}
and~\ref{table:2}. From these results, we have
\begin{align}
   \zeta_{11}(t)&=1+\frac{g_0^2}{4\pi}2(N-2)
   \left[\frac{1}{\epsilon}+\frac{1}{2}\ln(8\pi t)\right]+O(g_0^4),
\label{eq:(6.18)}
\\
   \zeta_{12}(t)&=\frac{g_0^2}{4\pi}(-1)(N-2)
   \left[\frac{1}{\epsilon}+\frac{1}{2}\ln(8\pi t)\right]+O(g_0^4),
\label{eq:(6.19)}
\end{align}
and from these
\begin{equation}
   \zeta_{22}(t)=\zeta_{11}(t)+(2-\epsilon)\zeta_{12}(t)
   =1+\frac{g_0^2}{4\pi}(N-2)+O(g_0^4).
\label{eq:(6.20)}
\end{equation}
Note that IR divergences are canceled out in the coefficients. Using these
results in~Eqs.~\eqref{eq:(6.13)} and~\eqref{eq:(6.14)}, we have
\begin{align}
   c_1(t)&=\frac{1}{g^2}-\frac{1}{4\pi}(N-2)\ln(8\pi\mu^2 t)+O(g^2),
\label{eq:(6.21)}
\\
   c_2(t)&=\frac{1}{4\pi}(N-2)+O(g^2)=b_0+O(g^2),
\label{eq:(6.22)}
\end{align}
where $g$ is the renormalized coupling in the $\text{MS}$
scheme~\eqref{eq:(3.5)}. Note that these coefficients are UV finite in terms of
the renormalized parameter. This must be so, because all composite operators
appearing in~Eq.~\eqref{eq:(6.12)} are renormalized ones.

For $c_2(t)$~\eqref{eq:(6.22)}, one may proceed one step
further~\cite{Suzuki:2013gza} by requiring that Eq.~\eqref{eq:(6.12)}
reproduces the trace anomaly~\eqref{eq:(6.5)} to the two-loop order. By taking
the trace of~Eq.~\eqref{eq:(6.3)} and comparing it with~Eq.~\eqref{eq:(6.5)},
we find
\begin{equation}
   \left\{\partial_\rho n^i\partial_\rho n^i\right\}_R(x)
   =\left[1+O(g^4)\right]\left[\partial_\rho n^i(x)\partial_\rho n^i(x)
   -\left\langle\partial_\rho n^i(x)\partial_\rho n^i(x)\right\rangle\right].
\label{eq:(6.23)}
\end{equation}
Using Eq.~\eqref{eq:(6.11)} with~Eq.~\eqref{eq:(6.20)}, we find that for
Eq.~\eqref{eq:(6.12)} to reproduce the trace anomaly~\eqref{eq:(6.5)} to the
two-loop order,
\begin{align}
   c_2(t)&=b_0+(b_1-b_0^2)g^2+O(g^4)
\notag\\
   &=\frac{1}{4\pi}(N-2)
   -\frac{1}{(4\pi)^2}(N-2)(N-4)g^2+O(g^4).
\label{eq:(6.24)}
\end{align}

The expression for the energy--momentum tensor that is usable with lattice
regularization is thus given by the $t\to0$ limit of~Eq.~\eqref{eq:(6.12)} with
the coefficients in~Eqs.~\eqref{eq:(6.21)} and~\eqref{eq:(6.24)}. As noted
above, we can replace the renormalization constant~$g$ and the renormalization
scale~$\mu$ in~Eqs.~\eqref{eq:(6.21)} and~\eqref{eq:(6.24)} by the running
coupling $\Bar{g}(q)$ with the renormalization scale~$q$ and set
$q=1/\sqrt{8t}$. We may use, e.g., the four-loop running
coupling~\cite{Beringer:1900zz},
\begin{align}
   \Bar{g}(q)^2
   &=\frac{1}{b_0\ell}
   \Biggl[1
   -\frac{b_1}{b_0^2}\frac{\ln\ell}{\ell}
   +\frac{b_1^2(\ln^2\ell-\ln\ell-1)+b_0b_2}{b_0^4\ell^2}
\notag\\
   &\qquad\qquad{}
   -\frac{b_1^3(\ln^3\ell-\frac{5}{2}\ln^2\ell-2\ln\ell+\frac{1}{2})
   +3b_0b_1b_2\ln\ell-\frac{1}{2}b_0^2b_3}{b_0^6\ell^3}
   \Biggr],\qquad
   \ell\equiv\ln\left(\frac{q^2}{\Lambda^2}\right),
\label{eq:(6.25)}
\end{align}
in actual numerical simulations.

\subsection{A facile computational method for~$\zeta_{IJ}(t)$}
In the above calculation of the matching coefficients~$\zeta_{IJ}(t)$, we have
regularized IR divergences by introducing the bare mass~$m_0$ for the
$N$-vector field. The required computation is, as a result, somewhat
troublesome. In this final subsection, we point out that, at least in the
one-loop level, there exists a ``facile method'' that avoids the introduction
of the IR-regularizing mass~\cite{Makino:2014taaa}. This method has been
particularly useful for gauge theories~\cite{Suzuki:2013gza,Makino:2014taa}
because one can regularize IR divergences without introducing a gauge-breaking
mass parameter; IR divergences are regularized by ``dimensional
regularization''.

We first note that for a Feynman diagram that contributes
to~Eq.~\eqref{eq:(6.16)}, e.g., the diagram in~Fig.~\ref{fig:15}, there always
exists a corresponding flow Feynman diagram that contributes
to~Eq.~\eqref{eq:(6.15)}. The topology of both diagrams is identical
(Fig.~\ref{fig:15} for the present example) but in the latter the propagators
carry the Gaussian damping factor~$e^{-t\ell^2}$, where $\ell$ is the loop
momentum, as in~Eq.~\eqref{eq:(2.15)}. As we have observed above, what is
relevant to~$\zeta_{IJ}(t)$ is the difference of the values of these two
diagrams which, by dimensional counting, has the structure
\begin{equation}
   \int_\ell\frac{\mathrm{e}^{-2t\ell^2}}{\ell^2+m_0^2}
   -\int_\ell\frac{1}{\ell^2+m_0^2}
   =\int_\ell\frac{\mathrm{e}^{-2t\ell^2}-1}{\ell^2+m_0^2}.
\label{eq:(6.26)}
\end{equation}
In this combination, IR divergences are canceled out and thus we may
set~$m_0\to0$ in this combination.\footnote{A flow Feynman diagram that does
not have its counterpart in the 2D field theory, such as the diagrams
in~Figs.~\ref{fig:13} or~\ref{fig:14}, is IR convergent; dimensional counting
shows that the loop integral has the structure
$\int_0^t\mathrm{d}s\,\int_\ell\mathrm{e}^{-s\ell^2}$ without the denominator.}
On the other hand, this integral contains UV divergences and we use the complex
dimension~$D$ to regularize this integral. For~$m_0\to0$, the result is given
by
\begin{equation}
   \int_\ell\frac{\mathrm{e}^{-2t\ell^2}-1}{\ell^2}
   =-\frac{1}{(4\pi)^{D/2}}\frac{1}{-D/2+1}(2t)^{-D/2+1},
\label{eq:(6.27)}
\end{equation}
as the analytic continuation from~$\re(D)<2$. This computation corresponds to
the computational method in the preceding subsection.

On the other hand, \emph{if\/} we forget to include the contribution
corresponding to~Eq.~\eqref{eq:(6.16)}, we will have only the first term
of~Eq.~\eqref{eq:(6.26)}:
\begin{equation}
   \int_\ell\frac{\mathrm{e}^{-2t\ell^2}}{\ell^2+m_0^2}.
\label{eq:(6.28)}
\end{equation}
This is UV convergent, but contains IR divergences for~$m_0\to0$. Thus, we set
$m_0\to0$ and instead use the complex dimension~$D$ to regularize
\emph{IR divergences}. The result is given by
\begin{equation}
   \int_\ell\frac{\mathrm{e}^{-2t\ell^2}}{\ell^2}
   =-\frac{1}{(4\pi)^{D/2}}\frac{1}{-D/2+1}(2t)^{-D/2+1},   
\label{eq:(6.29)}
\end{equation}
as the analytic continuation from~$\re(D)>2$.

Now, interestingly, the right-hand side of~Eq.~\eqref{eq:(6.27)} and that
of~Eq.~\eqref{eq:(6.29)} are identical as a function of~$D$. Thus, we may use
the latter method instead of the former. The latter is computationally much
simpler because only flow Feynman diagrams have to be computed and the IR
regulator~$m_0$ is not necessary. We have tabulated the result of this facile
method in~Table~\ref{table:3}. It can be confirmed that each entry coincides
with the difference between corresponding entries of~Tables~\ref{table:1}
and~\ref{table:2}, as must be the case; the resulting matching
coefficients~$\zeta_{IJ}(t)$ obtained directly from Table~\ref{table:3} are, of
course, identical to the previous ones,
Eqs.~\eqref{eq:(6.18)}--\eqref{eq:(6.20)}.

\begin{table}
\caption{The result of the facile method; in units of~$g_0^2/(4\pi)$.}
\label{table:3}
\begin{center}
\renewcommand{\arraystretch}{2.2}
\setlength{\tabcolsep}{20pt}
\begin{tabular}{crr}
\toprule
 diagram & \multicolumn{1}{c}{$ip_\mu iq_\nu+iq_\mu ip_\nu$}
 & \multicolumn{1}{c}{$2\delta_{\mu\nu}ip_\rho iq_\rho$} \\
\midrule
07  &
$-2\left[\dfrac{1}{\epsilon}+\dfrac{1}{2}\ln(8\pi t)\right]$ 
& $0$ \\
08  & $(2N-2)\left[\dfrac{1}{\epsilon}+\dfrac{1}{2}\ln(8\pi t)\right]$ & $0$ \\
09  & $\dfrac{3}{4}$ & $\dfrac{1}{2}N-\dfrac{5}{8}$ \\
10  & $-\dfrac{3}{4}$
& $(-N+2)\left[\dfrac{1}{\epsilon}+\dfrac{1}{2}\ln(8\pi t)\right]
-\dfrac{1}{2}N+\dfrac{5}{8}$ \\
\bottomrule
\end{tabular}
\end{center}
\end{table}

\section*{Acknowledgements}

We would like to thank Koji Harada for helpful discussion on perturbation
theory in the non-linear sigma model and Kengo Kikuchi for exposition on his
work, Ref.~\cite{Kikuchi:2014rla}. We are grateful to Martin L\"uscher for
explaining the hard part of the proof in~Ref.~\cite{Luscher:2011bx} patiently
to us. The work of H.~S. is supported in part by a Grant-in-Aid for Scientific
Research~23540330.

\section*{Note added in proof}
In recent papers \cite{Makino:2014cxa,Aoki:2014dxa}, the solution to the flow
equation~\eqref{eq:(2.2)} in the $1/N$ expansion is studied. In the former
work, the expectation value of~Eq.~\eqref{eq:(6.12)} at finite temperature is
computed for the large-$N$ limit and it is shown that the expectation value
correctly reproduces thermodynamic quantities in the presence of a
non-perturbative mass gap.

\end{document}